\documentclass{PoS}
\usepackage{slashed}

\title{Machines and Algorithms}

\ShortTitle{Machines and Algorithms}

\author{\speaker{Peter Boyle}%
      \thanks{Alan Turing Faculty Fellow}\\
      \thanks{Royal Society Wolfson Fellow}\\
      \thanks{Brookhaven National Laboratory}
      School of Physics, The University of Edinburgh, Edinburgh EH9 3JZ, UK\\
      E-mail: \email{paboyle@ph.ed.ac.uk}}

\abstract{I discuss the evolution of computer architectures with a focus on QCD and with reference to the interplay between architecture, engineering, data motion and algorithms. 
          New architectures are discussed and recent performance results are displayed.
          I also review recent progress in multilevel solver and integation algorithms.}

\FullConference{34th annual International Symposium on Lattice Field Theory\\
                24-30 July 2016\\
                University of Southampton, UK}

\begin{document}

\section{Introduction}

\begin{figure}[hbt]
\includegraphics[width=0.33\textwidth]{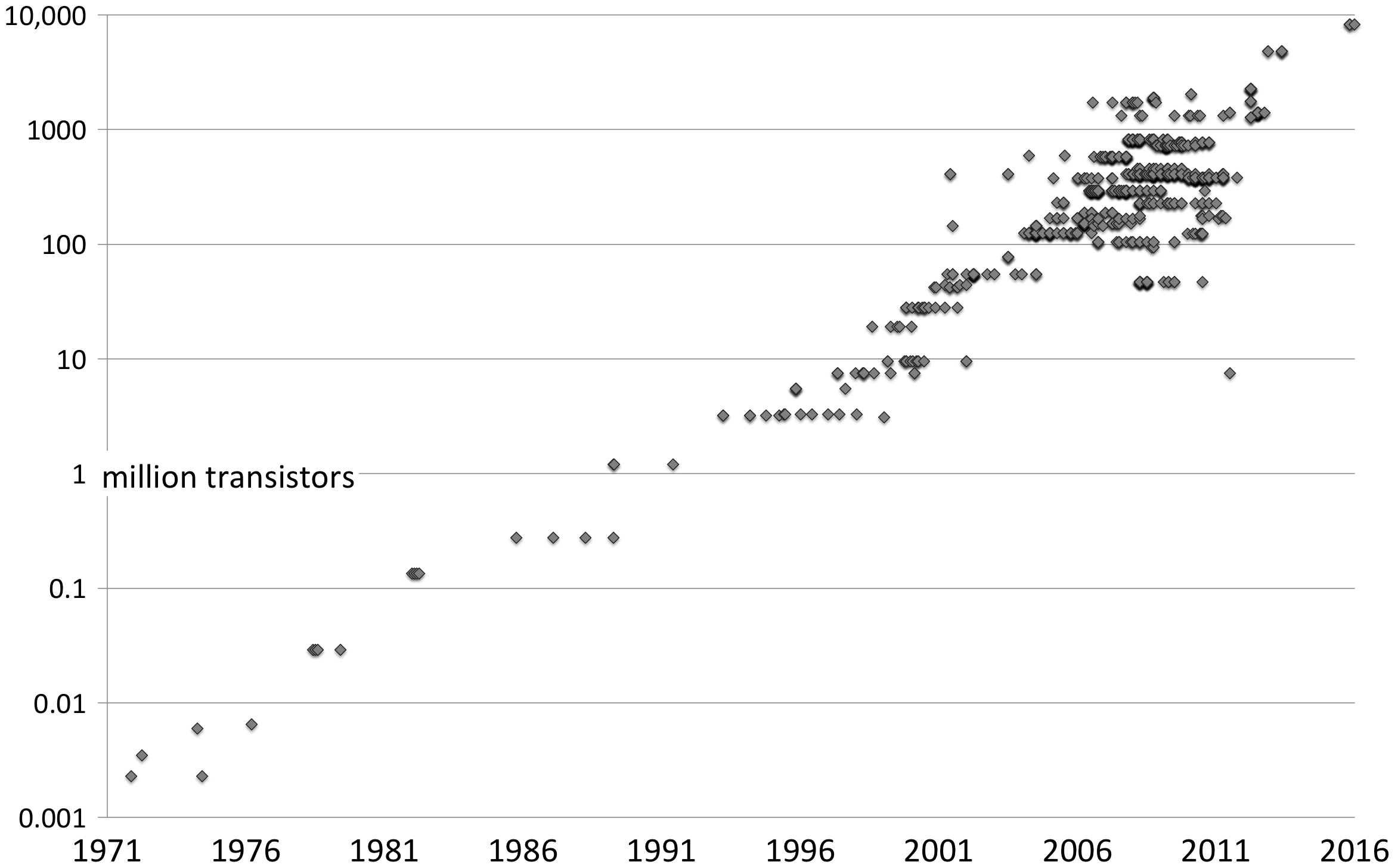}
\includegraphics[width=0.33\textwidth]{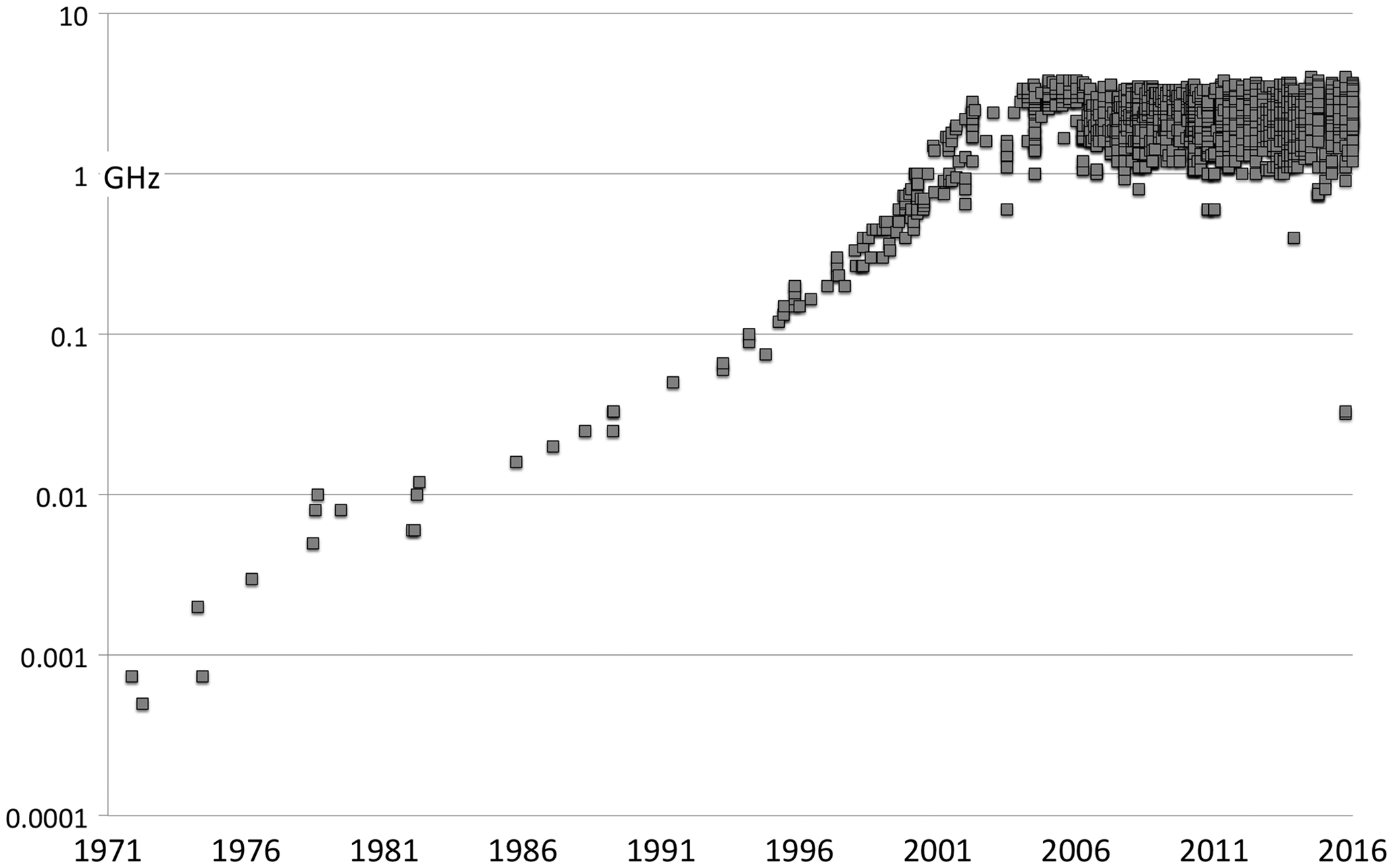}
\includegraphics[width=0.33\textwidth]{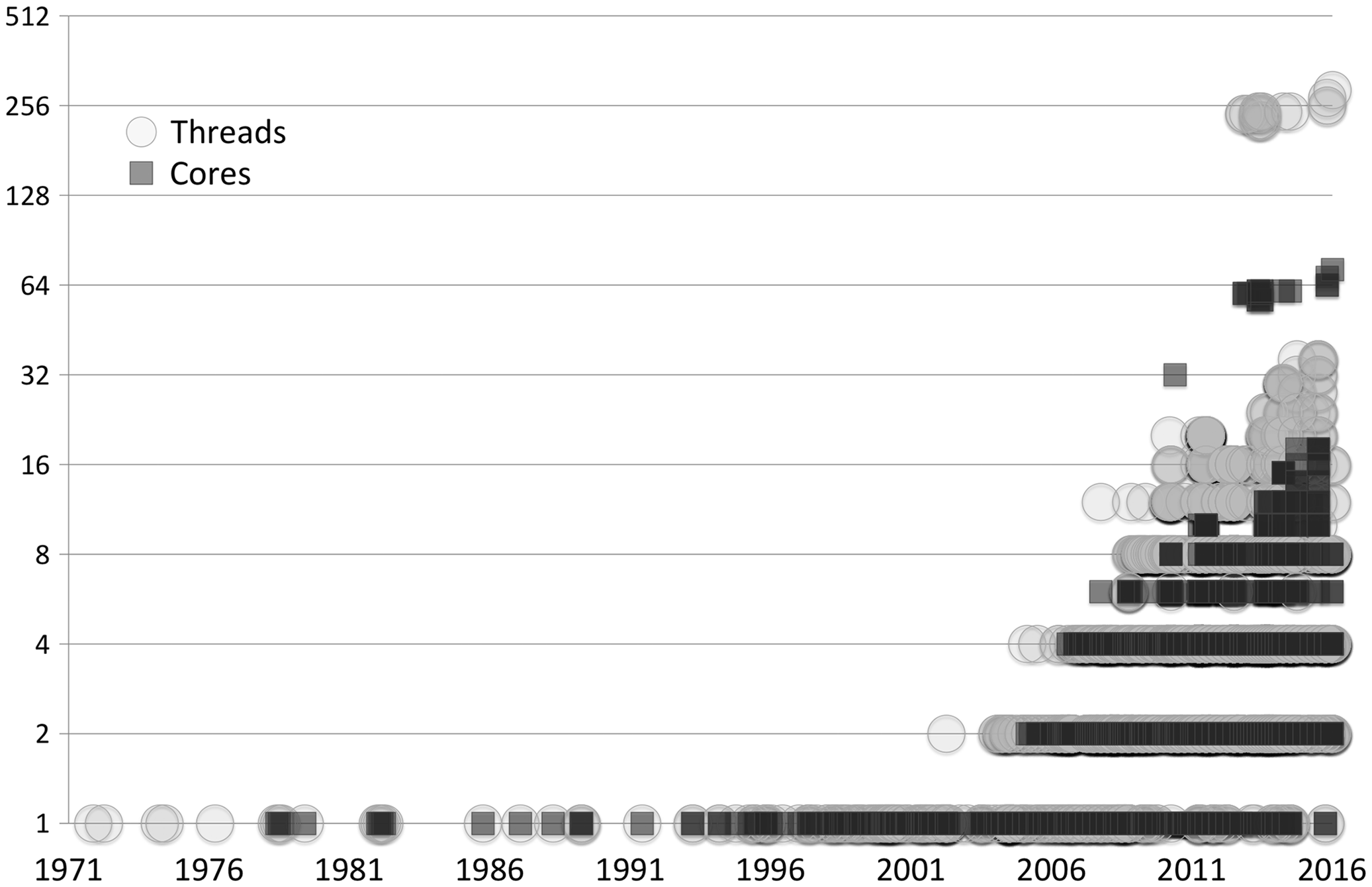}
\caption{ \label{fig:reinders} The continued Moore's law increase in transistor density has ceased to be accompanied
by growth in cpu frequency\cite{KNLbook,borkar}: the need to avoid transistor leakage has limited voltage reductions and gate delay 
has stalled; in addition a wire delay floor has been exposed which affects the degree to which a chip can be internally connected.
As a consequence the transister budgets have increasingly been spent growing on chip parallelism (cores, threads and vectors).}
\end{figure}

Present roadmaps for the evolution of computing paint a rather mixed picture for the scientific programmer: computer
power continues to grow rapidly. However this growth increasingly arises solely
through the expansion of many forms of parallelism, rather than improving serial execution speeds.
This is manifested, Figure~\ref{fig:reinders}, by a continued growth in transistor and hence core counts,
but no longer accompanied by greater clock frequencies. These figures are reproduced from ref~\cite{KNLbook}.

The sheer pain of efficiently programming increasingly complex devices is daunting, particularly
since the industry standard solution appears to be to dump it on the programmer.
The forms of parallelism include message passing interfaces for treating distributed memory multi-processing
(e.g. MPI), thread level parallelism (OpenMP, CUDA), and single-instruction-multiple-data (SIMD) short vector
instructions.
\begin{table}[hbt]
\begin{tabular}{ccccccc}
Core & simd &Year & Vector bits & SP flops/clock/core & cores & flops/clock\\
\hline
Pentium III& SSE    &1999 & 128 & 3 & 1 & 3\\
Pentium IV & SSE2   &2001 & 128 & 4 & 1 & 4\\
Core2        & SSE2/3/4 &2006 & 128 & 8 & 2 & 16\\
Nehalem      & SSE2/3/4 &2008 & 128 & 8 & 10 & 80\\
Sandybridge & AVX &2011  & 256 & 16 & 12 & 192\\
Haswell &AVX2     &2013 & 256 & 32 & 18 & 576\\
KNC & IMCI    &2012 & 512 & 32 & 64 & 2048\\
KNL & AVX512  &2016 & 512 & 64 & 72 & 4608\\
Skylake &AVX512 & 2017(?) & 512 & 96 & 28 & 2688
\end{tabular}
\caption{
\label{tab:chips}
We tabulate the floating point instructions per cycle provided on a range of commodity over the last two decades
to illustrate the increasing challenge placed on the programmer to coordinate this concurrent activity.
}
\end{table}

\begin{figure}[hbt]
\includegraphics[width=0.33\textwidth]{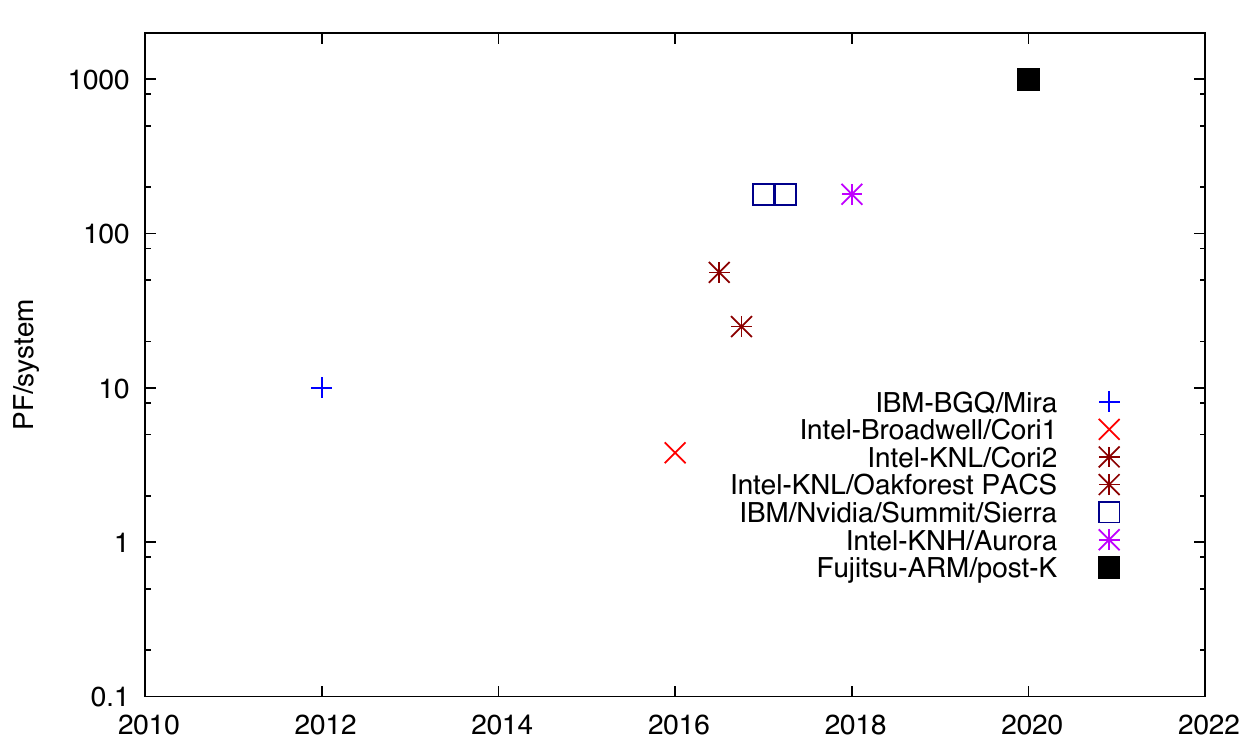}
\includegraphics[width=0.33\textwidth]{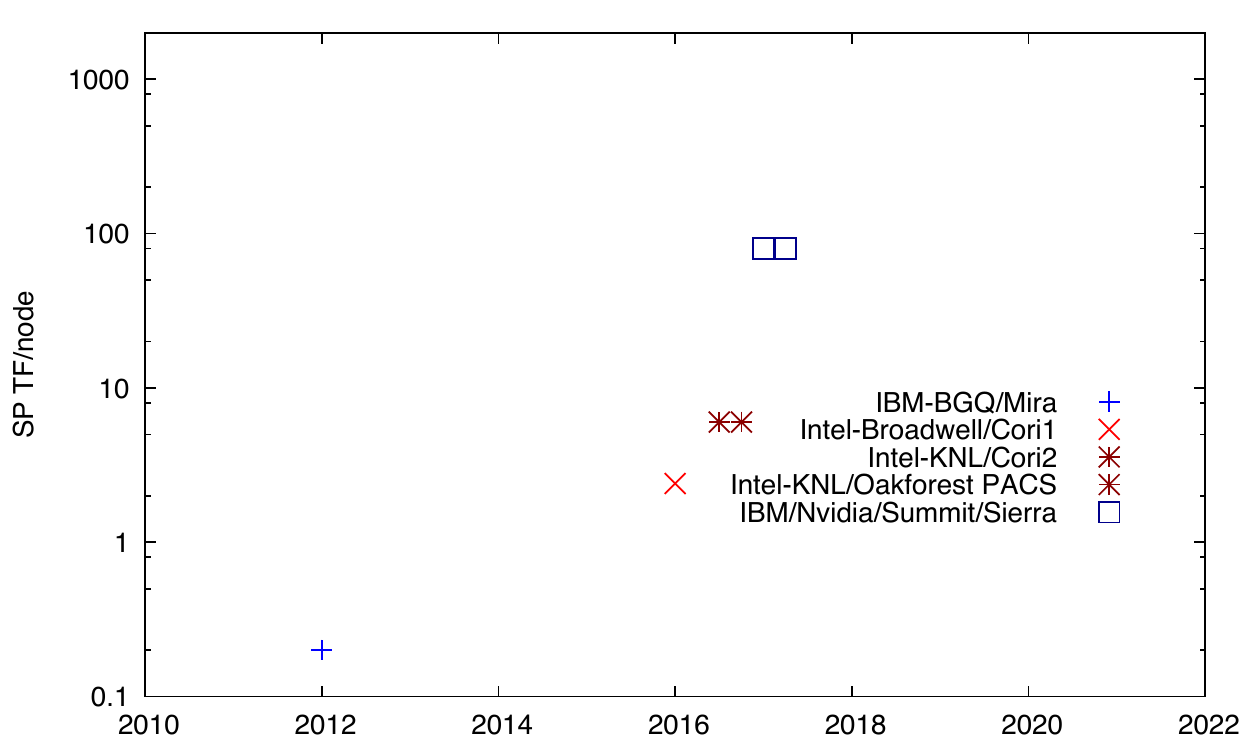}
\includegraphics[width=0.33\textwidth]{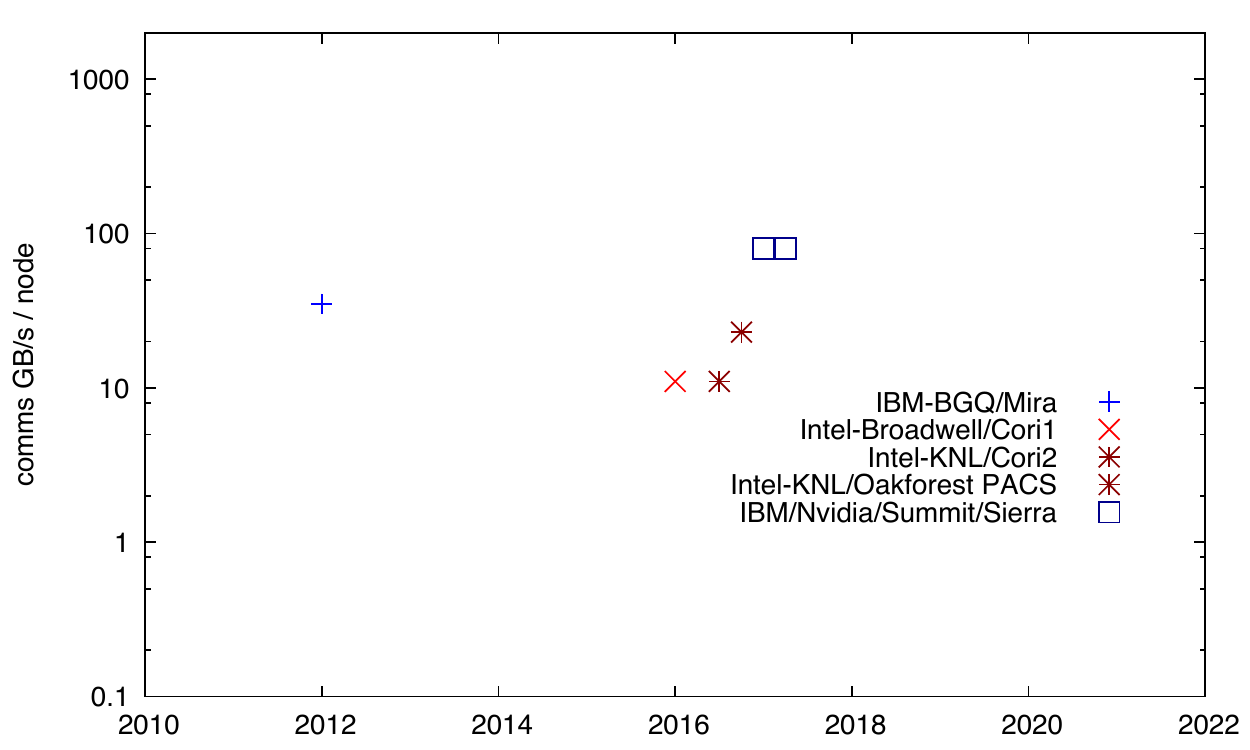}
\caption{ \label{fig:flopssys} Roadmap documented floating point peak per system (left), and per node (middle) versus planned
year of introduction. Increased computer performance largely comes from increased parallelism.
\label{fig:netbw} (right) Network bandwidth per node verus planned year of introduction.
Compared the BlueGene/Q system, which was highly scalable for QCD simulations a factor of 400 increase in per
node single precision performance anticipated, however this is accompanied by very little improvement in interconnect
bandwidth. It is clear that algorithmic changes will be required to make best use of such computers.}
\end{figure}

Further, in addition to complex but solvable programming challenges, Table~\ref{tab:chips}, there are fundamental limitations
on the flexibility of future systems imposed by changes in relative performance of floating point execution to the
bandwidth to caches, memory and interconnects. A timeline for the evolution of system performance, performance per node,
and interconnect performance is collated in Figures~\ref{fig:flopssys} and~\ref{fig:netbw}. We can see that
the four hundred fold project increase in single node performance in a few years is accompanied by only
a two fold increase in interconnect bandwidth. As a consequence, business as usual is not an option for algorithms.
We take as a starting point with machines such as today's BlueGene/Q system are very much in balance for scalable QCD simulation,
and plot the future evolution from published roadmaps 
of system arithmetic peak, single node arithmetic peak, and single node bidirectional network bandwidth in Figure~\ref{fig:flopssys}
and Figure~\ref{fig:netbw}.

\section{Technology directions}

I first discus the evolution and trends in the basic electronic engineering technologies that underpin
the development of computing. These represent the constraints within which vendors must operate.
On a point of principle it is worth emphasizing that our relations with vendors can be positive, but
we must understand each other clearly. Our interests are only
aligned when it comes to ensuring the best possible products.
Vendors have a requirement to maximise their profit, while
scientists have a requirement to maximise our scientific return, to a certain extent by minimising vendor profit.
Market segmentation techniques are the enemy of scientists, since these attempt to increase the price per core or Teraflop
as the aggregate capability per node or system increases. In short, we wish to perform our science at computer game pricing.

It is greatly to the benefit of science to ensure that the existence of a volume commodity market in similar
parts improves our science, rather than commercial profit margins while we buy the expensive version. This does imply that scientists
should always maintain a credible ability to use commodity parts. Our ability to do so is limited by reliability of low end 
``non server'' parts, and by the cost of efficiently interconnecting them. If the price differential is too great however,
we must bear in mind that we have the option of verifying solutions and/or recalculating a second time.

{\bf Wire delay:}
Some simple physics explains much of these computer architecture trends. We can model a wire as a rod of metal of volume
$L \times \pi r^2$. Applying Gauss's law,
$$
2\pi r L E = \frac{Q}{\epsilon},
$$
and the resistance and capacitance are
$$
R = \rho \frac{L}{\pi r^2} \quad\quad;\quad\quad
C = Q/V = 2\pi L \epsilon / log(r_0/r)
$$
This gives rise to the discharge time constant
$$
RC = 2 \rho \epsilon \frac{L^2}{r^2} / \log(r_0/r) \sim \frac{L^2}{r^2}
$$
This is profound, because to a great degree (ignoring logarithmic terms and finite size effects)
wire delay depends \emph{only} on the geometry or aspect ratio of the wire. Sadly, shrinking 
a process by the same scale factor in all dimensions does not speed up wire delay. Wire delay places
an irreducible floor below which faster transistors make no difference. It is interesting to note
that historically process advances announced in the press have included materials changes designed
solely to move this wire delay floor out of the way of transistor scaling:
``copper interconnect'' (180nm) and ``low-k`` dielectric (100nm) improved $\rho$ and $\epsilon$ respectively
changing the RC timeconstant. Anecdotally, the author participated in the design of memory prefetch engine
for BlueGene/Q in IBM's 45nm process. The slowest timing path in this component was roughly 50\% wire delay
and 50\% transistor delay; this balance becomes worse as transistors become faster without changes to the
materials that control wire delay.

The immediate corollary of the aspect ratio dominance of wire delay is the following:
\emph{Multi-core design with long-haul buses only possible strategy for 8 Billion transistor devices}.
There must be a low number of long range ``broad'' wires (bus/interconnect) since these consume area,
and a high number of short range ``thin'' wires giving densely connected regions (cores).

{\bf Memory technologies:}
Data motion is often the single greatest performance and power 
bottleneck to be addressed in computation. A nuanced approach to analysing code
will distinguish data references by their most likely origin (e.g. cache, local memory, remote node).
For a bandwidth analysis, the ratio of the floating point instructions executed to the bytes accessed 
for each of the cache, memory and interconnect subsystems is called the \emph{arithmetic intensity} of the
algorithm and is a key classfication to understand if a computer is engineered to support efficient execution
of the algorithm.
Since the cost of floating point units has decreased exponentially, while the cost of macroscopic
copper traces has remained (relatively) static, it is often the case that execution is memory limited.
The term memory wall has been coined representing a latency bound on execution throughput\cite{mwall},
and can impose a bound on throughput that can only be addressed by changes to the speed of light. Fortunately,
for deterministic algorithms, hardware or software prefetching can move the problem to a bandwidth problem
that succumbs to investment, and many scientific algorithms can be categorised as either bandwidth or FPU
bound, for example using the Berkely roofline model\cite{roofline}.

One important application of the previous simple analysis of wire delay arises when we consider connection of a
calculation device to memory chips. Long copper traces across a printed circuit board fall into the precisely
the worst case corner for wire delay. Further, given the speed of light and GHz scale clock rates 
distances of order 30cm (= 1 light-nanosecond) set the length scale at which transmission line behaviour
sets in. 

\begin{figure}[hbt]
\includegraphics[width=0.3\textwidth]{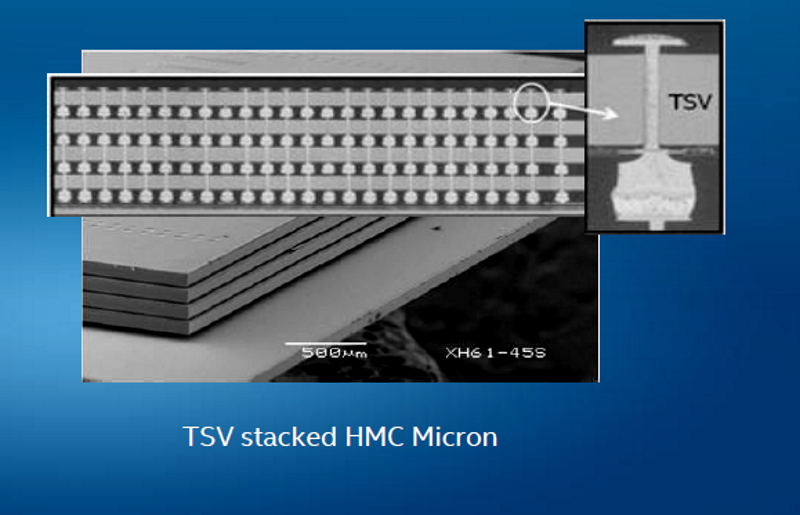}\hspace{0.2cm}
\includegraphics[width=0.3\textwidth]{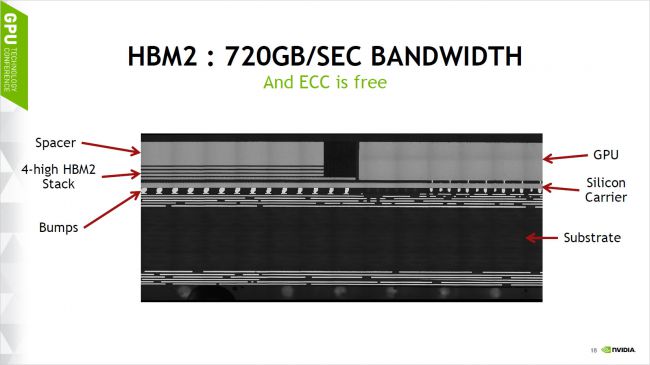}\hspace{0.2cm}
\includegraphics[width=0.3\textwidth]{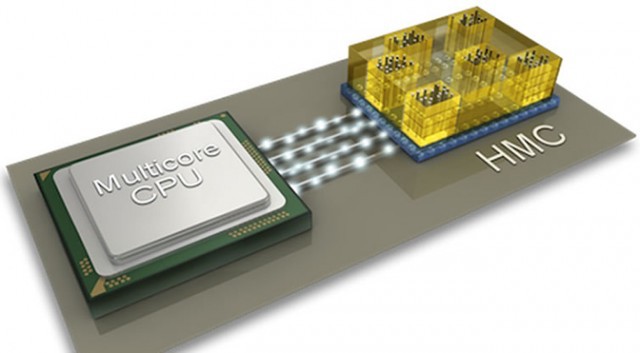}
\caption{ \label{fig:TSV} (Left) We display a micrograph of a through silicon via bus structure giving thousands of bit lanes
connecting memory chips with using a stacked configuration to give wiring geometries favourable from an energy and delay perspective.
Examples are displayed in use in recent 2.5D computational devices from Nvidia Pascal GP100 (HBM with silicon interposer)\cite{gp100}, and Intel Knight's
Landing (using Intel's own memory approach) \cite{KNLbook}.
}
\end{figure}

In order to substantially address memory limitations it is necessary to both change the aspect ratio of the wires
involved, and vastly increase the number of wires that can carry data concurrently.
Fortunately the computing industry has already invested in technology directions responding to this imperative;
the most significant standards are the ``High Bandwidth Memory'' (HBM), ``Hybrid Memory Cube'' (HMC), and ``WideIO''
as well as some proprietary interfaces.
Memory chips have been developed with large numbers of vertical rods (through silicon vias or TSV's) used as high bit count buses skewering 
the device, Figure~\ref{fig:TSV}. The ability to stack such chips vertically, with solder connections between TSV's giving a favourable aspect ratio,
low wire delay and power. Further the miniaturisation enables a high lane count connection that gives a massive data rate for reading memory pages;
it is exciting that after decades of conservatism in memory design, there is no clearly defined barrier to further increases in the number of bitlanes.

This is driving a rather revolutionary development across the industry towards on package integration of memory,
where a sizable volume of memory is physically colocated with the processing unit.
This may be done either through 2.5D integration with memory stacks placed along side a processing unit
on an interposer (i.e. a miniature circuit board), or full 3D integration where memory chips are bonded on-top
of a processing unit using TSV's.

The HMC and HBM solutions are presently 2.5D and either exposed with high bit count and (relatively) low data rate per bit lane 
(HBM) to a memory controller on a processing unit, or in the case of HMC interfaces through a base layer logic chip that 
provides a high bit rate and low lane count interface to the processing unit.
Full 3D memory solutions could, in principle, yield significant growth of bus widths in future, since the mechanical assembly
for high density TSV's is fundamentaly different to the decades old PCB copper trace problem. 

A further significant emerging memory technology is a novel form of non-volatile phase change memory, typically based
on an amorphous/crystaline glass cell. This should enable to increase memory density, and in particular
is suggested to support many bit-cells vertically stacked in multiple layers. Micron and Intel place
this under ``3D Xpoint'' branding and promise four times higher density that DRAM. The most exciting application lies in plans
to give such devices conventional DRAM electronic interfaces, but uses will range from solid state disks to 
potential integration in conventional memory systems. Although details are rather
difficult to come by\cite{WikiPediaXpoint}. The author expects that it is a reasonable assumption that once a page is open the column accesses
should be no slower than conventional DRAM since the electronics will be similar, while page miss latency 
may be larger. Certainly there are 
already multiple JEDEC NVDIMM standards planned, and NVDIMM is likely 
a disruptive technology for large memory algorithms, such as the use of Dirac eigenvectors and the assembly
of multi-hadron correlation functions.

{\bf Interconnect:}
A very significant trend is the integration of interconnect.
Historically, QCD calculations have often been near the frontier of interconnect technology requirements,
largely because a four dimensional application with the same data footprint per node suffers from worse
surface to volume ratio than lower dimensional applicaton. As a result, integration of interconnect has
been pioneered by QCD machines such as QCDSP and QCDOC\cite{QCDSP}, and also by the IBM BlueGene
line\cite{BlueGene}. If the integration is performed on a compute ASIC, the additional cost for interconnect can
be reduced from thousands of dollars to pennies per node particularly if novel rack integration keeps
most traces in copper backplanes and driven by integrated transceiver logic.

There are some interesting current trends. Firstly, with the Intel Omnipath interconnect being integrated
\emph{on package} in one variant of the Intel Knights Landing device (KNL-F) and with variants of the
Intel Skylake Xeon processor. These use a distinct silicon chip implementation providing two 100Gb/s Omnipath
ports, and in the KNL-F is integrated in a multi-chip module package. The marginal cost is claimed to be around \$300 per node,
but this includes only the cost of the interfaces does not include cable and switch pricing. Further, a second generation
Omnipath 2.0 will be integrated with the Knights Hill computer chip planned for the Aurora supercomputer in 2018.
The principal competing interconnect technologies are the Mellanox Infiniband product line, which offers excellent performance but
typically with a greater per port cost, and the Cray Aries interconnect used in the Cori and Theta supercomputers as an example.

Secondly, Nvidia has announced a novel NVLink technology with a tremendous bandwidth of up to 
160GB/s bidirectional per link that scales up 8 GPU's. However, while providing excellent
local bandwidth within a server, this is \emph{not} a large cluster interconnect, and must be combined
with another technology in a large system. For example, the Summit and Sierra systems combining IBM Power hosts
with multiple Nvidia GPU's will make use of Mellanox interconnects between powerful compute.

{\bf System design and reduced optical cost:}
System or rack level design can make a large difference to interconnection costs.
Due to the attenuation of high speed signals in copper traces, it is unfortunately the case that 
almost all high performance transmissions over 3m in length must proceed using optical interconnects.
A significant design element of the BlueGene/Q system was its high density; by using very dense
racking techniques the optical cost was suppressed by the surface to volume ratio. Similar approaches
are taken to varying degrees in the Cray Aries and SGI ICE-X architectures, and in more standard blocking
fat tree networks.

\begin{figure}
\includegraphics[width=0.4\textwidth]{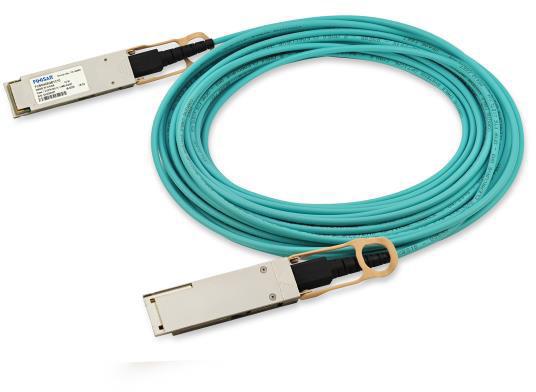} 
\includegraphics[width=0.4\textwidth]{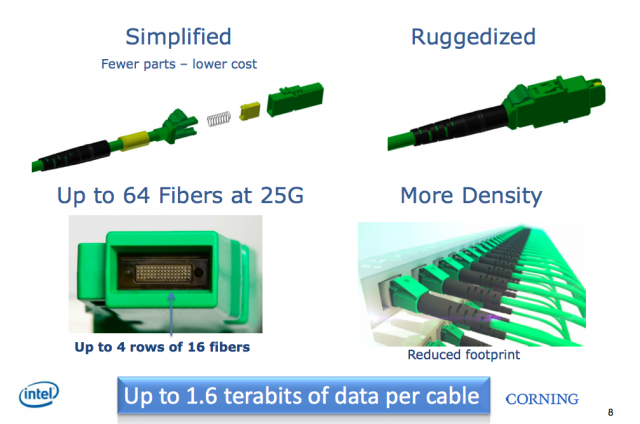} 
\caption{ \label{fig:figAOC} Left: an active optical cable carrying just four bit lanes in each direction and costing around \$1000 USD.
The cost is dominated by the transceiver electronics on either end.
Right: a passive optical cable carrying 64 bit lanes. There is little prospect for an further improvement in density; the pitch is
set by the need for a single grain of dust to not block the light path of any bit. Indeed, optical lenses broaden the beam inside these connectors
to enhance the blockage tolerance. }
\end{figure}

While 100Gbit/s of copper trace in a larger PCB will cost as little as pennies, the same
100Gbit/s of bandwidth in copper cables costs just under \$100 USD. A similarly performing 100Gbit/s 
active optical cable with four bit lanes costs an eye watering 1000 USD\ref{fig:figAOC}. It may even be the case that
never before in the field of computing have so few bits cost so much!
This  expense is largely due to the use of copper signalling and active optical transceivers on either end of the cable.
The cost is largely in the electronics, and may in future be addressed if \emph{silicon photonics} allows the integration
of the laser transceivers upstream in the computational node.
This has the potential to reduce the cost and power of driving fibre cable to be closer to cost of copper, and to use a normal silicon process for laser components.

There is probably a room for an adventurous vendor or even academic project
to design a standard cluster but with dense rack packaging and backplane local interconnect,
perhaps even using PCI express at the rack level to suppress the cost of interconnect, in a similar style to the BlueGene/Q rack level design,
but otherwise using commodity components.

\section{Novel hardware platforms}

The two significant hardware platforms released in 2016 are the Intel Knights Landing ``many-core'' processor
and the  Nvidia Pascal GPU. These give much more parallelism than more conventional nodes from Intel and IBM,
and greater power efficiency, and there is considerable community interest in whether the greater peak
performance and efficiency translates into delivered application performance for QCD. We survey single
node performance results from QCD software development efforts on these chips.

\begin{figure}[hbt]
\includegraphics[width=0.4\textwidth]{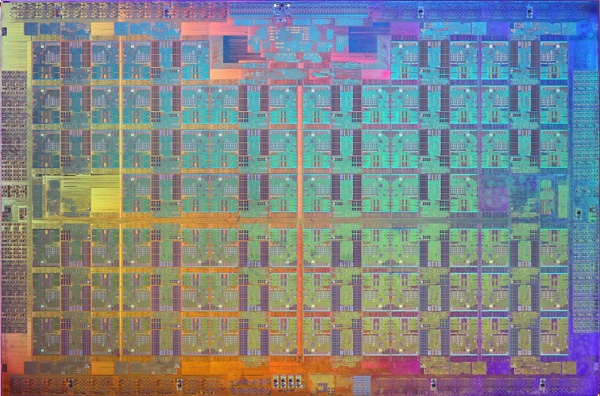}
\includegraphics[width=0.3\textwidth]{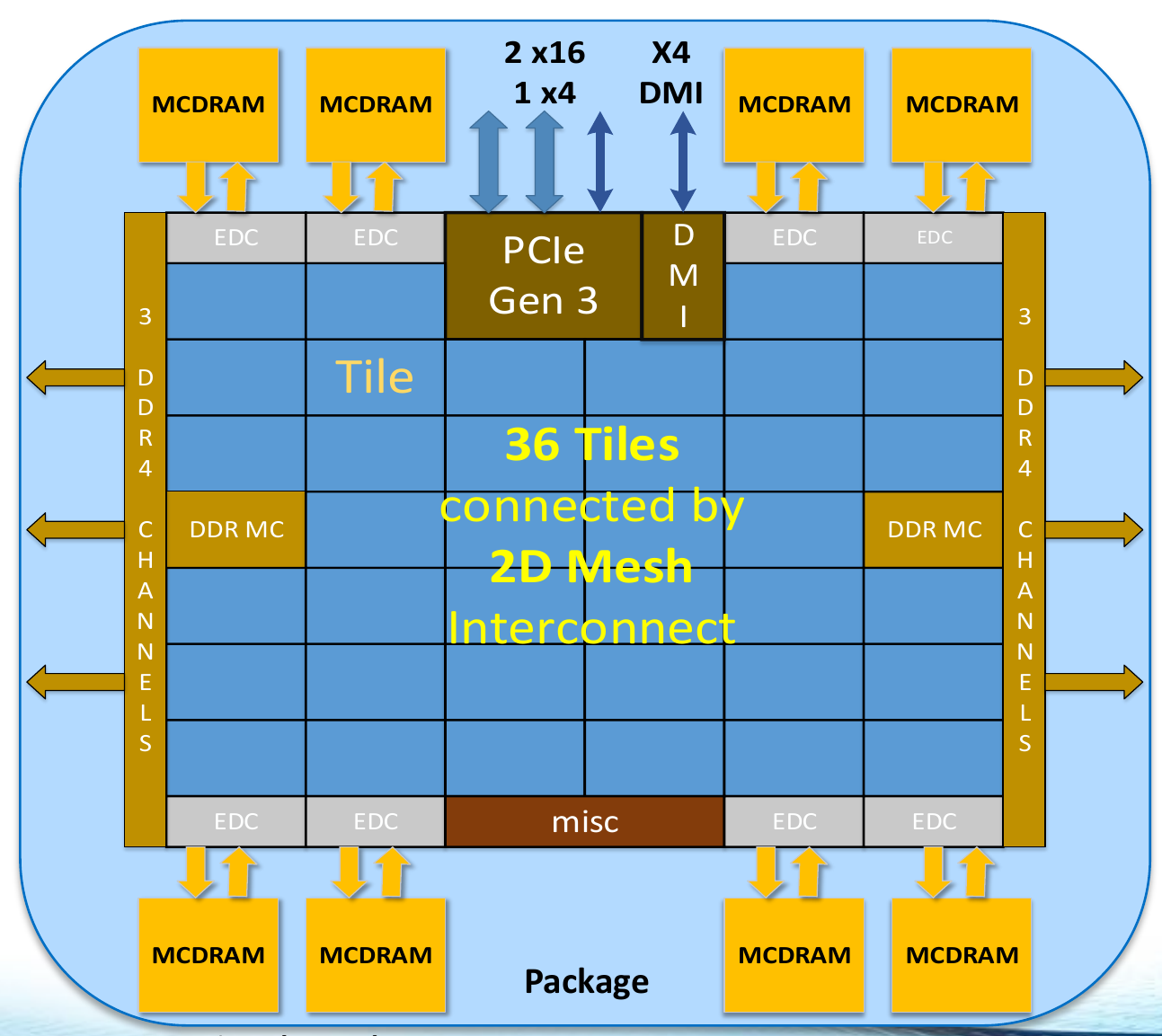}
\includegraphics[width=0.25\textwidth]{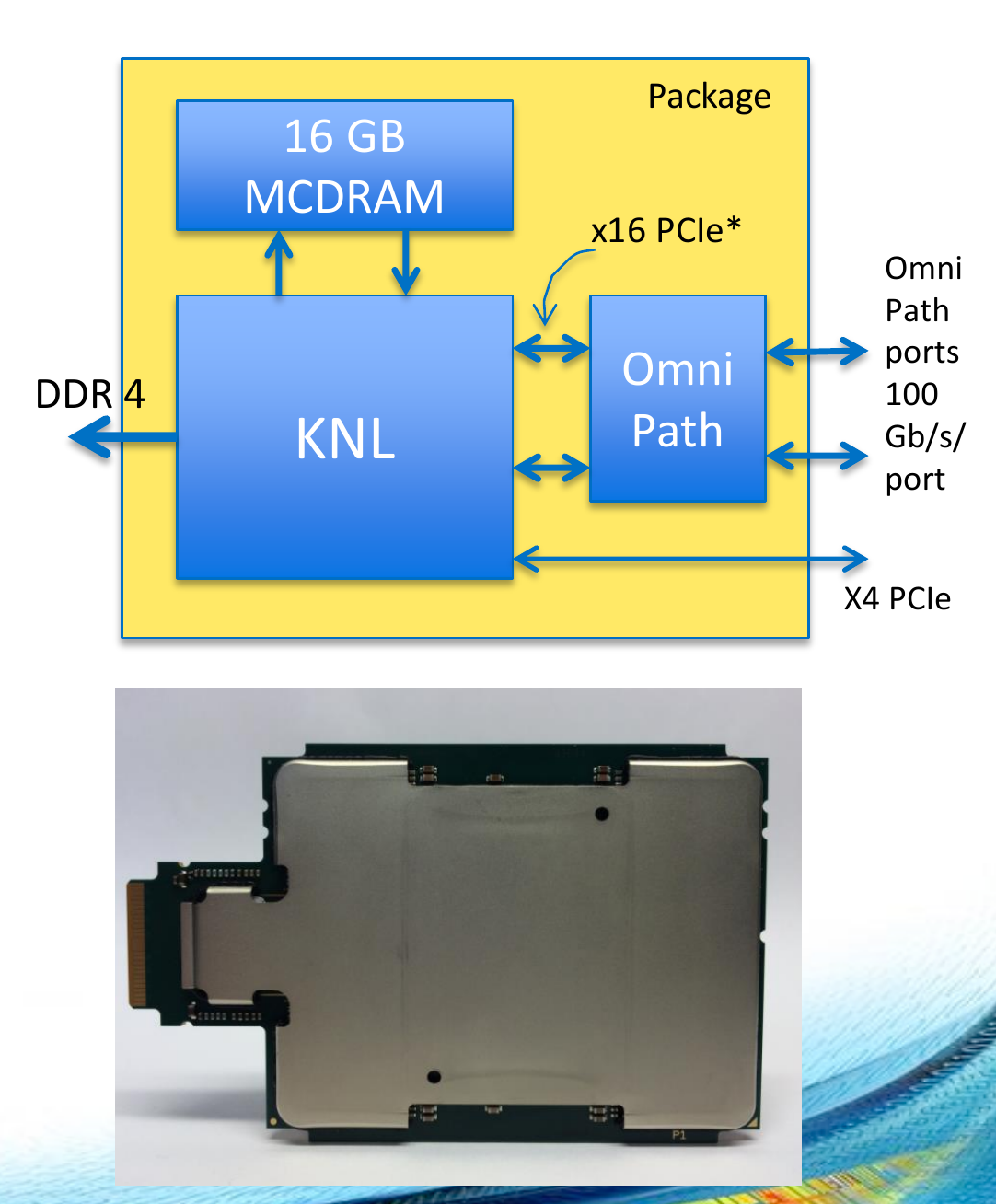}
\caption{\label{fig:knl}
The Intel Knights Landing processor die and a schematic diagram of the on-chip tiled arrangment and memory
system mesh. High bandwidth memory on package augments the 6 channel DDR4 interface giving good memory bandwidth.
36 lanes of PCIe Gen 3.0 interface is included, in principle giving a good I/O bandwidth (over 64GB/s bidirectional) to whatever
interconnect one can afford to combine this node with. The Tile consists of a pair of processor cores and 1MB local L2 cache.
}
\end{figure}

{\bf Intel Knights Landing:}
Intel has introduced the Knights Landing chip\cite{sodani} makes use of many low power, but architecturally standard, 
processor cores. The cores can issue up to two instructions per cycle (compared to up to 6 instructions per cycle for high end Xeon cores)
and this includes up to two 512 bit wide vector instructions. The devices range from 64 to 72 cores depending
on the specific part. The cores execute all legacy x86 instructions such as SSE, AVX and AVX2 but best performance
requires targetting the AVX512 instruction set which can be a non-trivial software investment. The peak performance using
single precision is over 6 TFlop/s and over 3TFlop/s for double precision.

The cores are arrange into tiles each containing a 1MB L2 cache and a pair of cores. These
are interconnected through a two dimensional routing mesh, and the system-on-a-chip includes both
six DDR4 memory controllers providing up to 384GB of memory
(quoted at 90GB/s STREAMS AXPY) and eight on-package chipstacked memory controllers (quoted at 400GB/s STREAMS AXPY) 
providing 16GB of memory.
The high bandwidth memory can either be used in \emph{flat} mode as a directly addressable NUMA domain under Linux, in \emph{cache} mode
as a very large direct mapped L3 cache, or in \emph{hybrid} mode as 8GB NUMA domain and 8GB cache.

The Knights Landing can be packaged with two ports of 100Gb/s Omnipath interconnect fabric, and unlike the previous
Knights Corner chip can be configured as a self hosting compute node, without the need of an accelerator card or offload
programming model. This is a substantial simplification for the programmer.
A sensible way to programme these chips involves using a hybrid combination of MPI message passing, OpenMP threading and
SIMD vectorisation. The nature of modern compilers and some experience suggests that the most effective vector utilisation
is obtained when compiler vector intrinsic functions are used, rather than relying on compiler vectorisation of loops.

\begin{figure}[hbt]
\includegraphics[width=.45\textwidth]{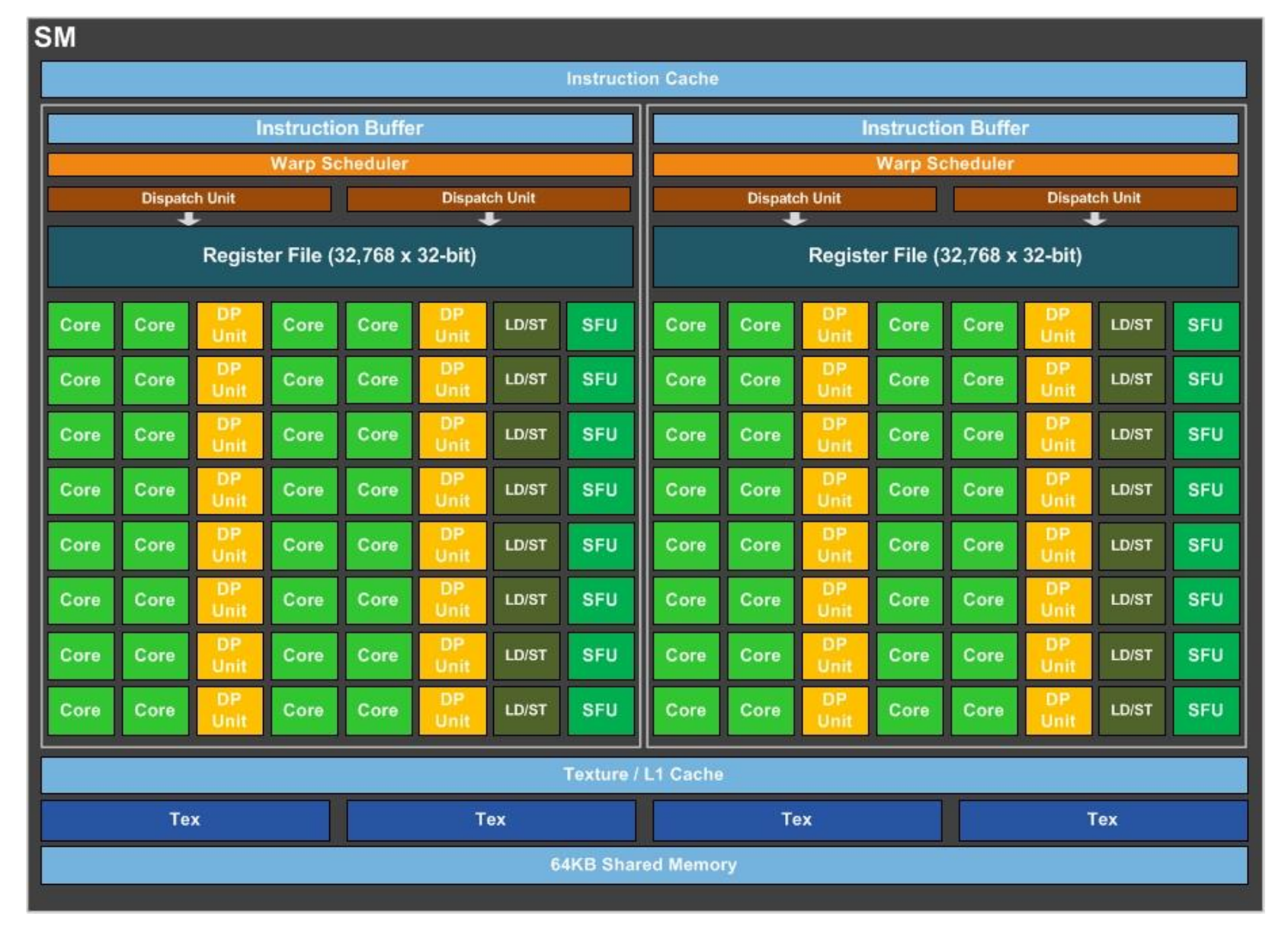}
\includegraphics[width=0.49\textwidth]{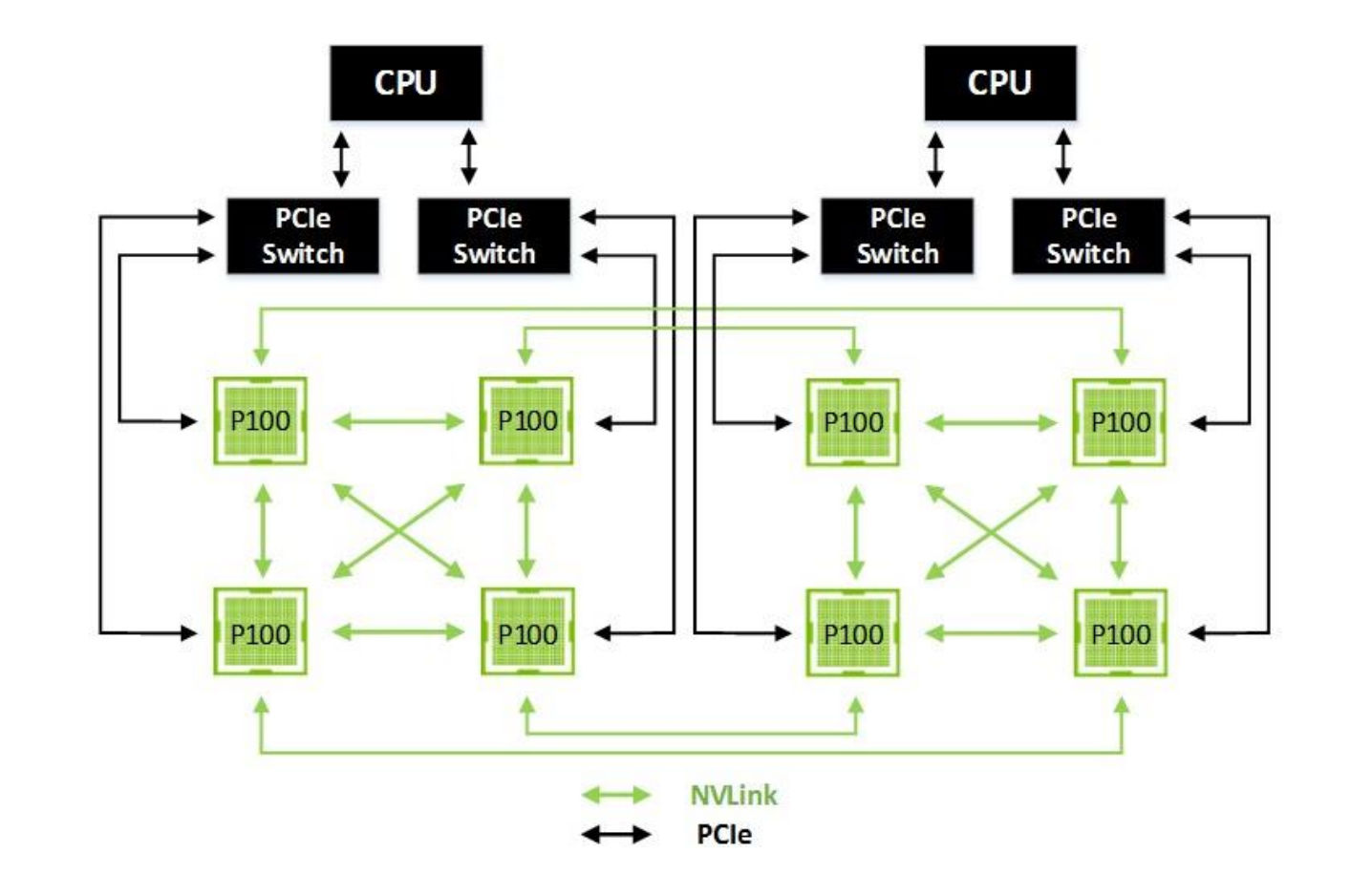}
\caption{\label{fig:pascal}
Left: The Nvidia PASCAL GPU consists of a number of ``streaming multiprocessors''. Each of these has a
single instruction fetch engine which can operate in a data parallel fashion on a number of SIMD lanes.
These SIMD registers contain both floating point and integer ``programme state''. Nvidia refers to 
each lane as a CUDA core, however the CUDA cores do not fetch instructions independently. If different
threads branch in a data dependent manner the shared instruction fetch must sequence both directions serially.
Given the difference in the nature of the core between architectures, it is most reasonable to compare
the number of floating point SIMD lanes between CPU's and GPU's as a like for like comparison.
Right: The new NVLink interface demonstrating internal node connectivity for multi-GPU nodes. In some IBM Power8 products NVLink can replace
PCI express for the purpose of bulk transfers to the host memory and/or interconnection network helping to address
bottlenecks in scaling multi-GPU simulations on previous generations.
}
\end{figure}

{\bf Nvidia Pascal:}
Nvidia has released the Pascal GPU, with some exciting features and impressive peak single precision performance of 10.6 TF/s,
Figure~\ref{fig:pascal}. The Pascal uses a form of 3D memory known as HBM2, providing up to 600GB/s Streams AXPY\footnote{Figure
provided by Kate Clark.}. The device supports 64 bit, 32 bit, and 16bit floating point arithmetic, and 
various reliable update, preconditioning, or mixed precision schemes may be used to accelerate
QCD solution without loss of accuracy. 
The Pascal introduces 4 NVLink ports, Figure~\ref{fig:pascal}, per GPU giving up to 160GB/s of bidirectional bandwidth in aggregate
across these links. The NVLink ports can be used to interconnect up to eight GPU's within an Nvidia DGX1 system, and also
can accelerate the bottleneck of a slow PCIe I/O bus between up to four GPUs and the host memory (and MPI interconnection)
when combined with IBM Power8 nodes in IBM S822LC servers. 

In common with earlier GPU versions, the simple cores are only accessible through an offload
programming model since they cannot run general code. These are typically accessed through the
CUDA programming environment that leverages the C++ language. This has been made accessible
for QCD calculations by both the QUDA library \cite{quda} and the QDP-JIT library\cite{qdp-jit}.
Single nodes with up to 40 or 80 peak single precision TF/s can be assembled using 4 or 8 GPUs. 
For many QCD algorithms interconnect performance becomes a key consideration, and communication minimising strategies
such as domain decomposition will be important.

In contrast to earlier GPU's, Pascal introduces a ``Page Migration Engine'' that can
transparently move virtual memory pages between the CPU host and the accelerator card.
This introduces a common address space, even if implemented through physical motion, between code
executing on the host and on the accelerator. This is potentially a programming model simplification, allowing
a single allocator and no explicit data motion. There is great scope for simpler QCD libraries
to make use of Pascal and future Nvidia GPU's leading to more easily performance portable code.

A sensible way to programme systems based on these chips involves using a combination of MPI message passing and CUDA calls.
The CUDA code still requires data to be laid out in a SIMD or vector friendly order in order to ensure that
single memory transactions are performed for all elements of a SIMD unit; this optimisation is known as read coalescing.
As a result, both GPU and many-core CPU code requires layout flexibility to be optimal; however at present it is rather
difficult to write computer programmes that support both with best efficiency.

\section{Software challenges}

Text book computer engineering\cite{HennessyPatterson} suggests that code optimisations should
expose an admixture of two forms of locality: spatial data reference locality, and temporal data reference locality.

Buses are used that transfer many contiguous words of data at a time in many levels of a computer memory system.
For example, When a DRAM page is opened (row access), very many consecutive bit cells are read and \emph{page hits} (column access) incur much less 
overhead than a new random access giving rise to different latencies. Further, multiple layers of caches are used; the access bandwidth and latency
characteristics become poorer the further one travels from the processor core while the storage capacity correspondingly increases.
The transfers between levels in the memory system are performed in aligned chunks called cachelines, with sizes $\in\{32,64,128,256,512\}$ bytes
depending on the level and computer architecture. For example, modern Intel chips typically use a 64B cacheline in all layers of the cache hierarchy, while
the IBM BlueGene/Q system made use of a 64B L1 line size and a 128B L2 line size.

Spatial reference locality arises because memory systems are fundamentally \emph{granular}. If one touches and transfers one word
of a cacheline, the entire cacheline is transferred over (rate limiting) buses in the system. Algorithms and codes should ensure that they 
make use of all the data in cachelines that are transferred (by laying out data to give spatial locality of reference) to maximise performance.
Architectures use cachelines based on the observation that many algorithms (such as linear algebra) have such access patterns.
Temporal locality of reference arises when data referenced is likely to be reference again soon. Thus caches store recently accessed
data.
Both of these hardware optimisations can be exploited in software by writing scalar loops that loop or block in the right order, and so are
relatively easy to exploit. 

In contrast, short vector instructions (single instruction, multiple data or SIMD) brings a new level of restrictivness 
that is much harder to hit. Code optimisations should expose what I will call spatial operation locality.
Although there are obvious applications in array and matrix processing, even matrix transposition shows that it is surprisingly hard to exploit 
in general because you must both arrange to have same operation applied to consecutive elements of data.

However, it is vastly cheaper to build hardware execute a single arithmetic instruction or load on a contiguous block of $N$ elements of data than to generally
schedule $N$ independent and decoupled instructions. As a result, SIMD is one of the cheapest ways to enhance peak floating point performance,
particular if the rest of the processor core is rather complex and aggressively optimised. Table~\ref{tab:simd} displays the decomposition of various
current computational chips into processing cores and SIMD execution units.

\begin{table}[hbt]
\begin{tabular}{c|c|c|c|c|c|c|c}
Chip & Clock & blocks & per bock  & SP madd    & issue           & SP madd & peak\\
\hline
GP100 & 1.4 GHz &  56 SM's     & 2 IB/RF & 32 & $112 \times 2$  & 3584 & 10.5 TF/s\\
KNL   &1.4 GHz &  36 L2 tiles  & 2 cores & 32 & $72\times 2$    & 2304 & 6.4 TF/s\\
Broadwell & 2.5 &  & 18 cores & 16 & $18\times 2$   & 576  & 1.4 TF/s
\end{tabular}
\caption{\label{tab:simd}
The decomposition of various
current computational chips into processing cores and SIMDexecution units. It is worth noting that nomenclature varies between different
vendors, and itcan even be difficult to match languagedue to architectural differences. Nvidia calls each SIMD lane a core despite instruction
fetch and issue logic being common; this is forgood reason because each SIMD lane can support a completely distinct programme state or``thread''.
However, it is probably most reasonable to compare the CUDA core count quoted for GPU'sto the SIMD lane count in CPU's since this counts the total
number of floating point pipelines in both cases.
 }
\end{table}

\section{QCD computational challenge}

We will consider the computational requirements of Wilson, Improved Staggered and Domain Wall fermion actions and with both 
single right hand side and multiple right hand side Krylov solvers. 
Multiple right hand side solvers can be advantageous, where applicable, for Wilson and Staggered 
actions because the gauge field is a significant element of the memory traffic and can be applied to multiple
vectors concurrently, suppressing this overhead. 
The same suppression factor already occurs quite naturally for five dimensional chiral fermion actions such as DWF.
When this is combined with \emph{block solvers}\cite{Wagner}, this conference, the gain is amplified and both algorithmic and execution
performance can be accelerated.
We assume an $L^4$ local volume and either an  8 point (nearest neighbour) or 16 point (nearest neighbour + naik) stencil.
Providing the global switching is implemented well in a fat tree switched network, or we map well to the topology of
a routing torus, hypercube, or Dragonfly network this will suffice to characterise the performance of a large
machine. That is to say under the assumption weak scaling, the local floating, memory and cache bandwidths and MPI network bandwidth per node
suffice to describe larger systems. This assumption may be invalidated if networs implementation is poor. 

For either multiple RHS or for DWF we  take $L_s \equiv N_{\rm rhs}$.
A cache reuse factor $\times N_{\rm stencil}$ on Fermion is possible, but only if the cache is of sufficient capacity.
We can count the words accessed per 4d site of result by each node, and the surface volume that must be communicated
in a massively parallel simulation with this local volume. These comprise
\begin{itemize}
\itemsep-0.3em 
\item Fermion: $N_{\rm stencil} \times (N_s \in \{ 1,4 \}) \times (N_c=3) \times (N_{\rm rhs} \in \{ 1, 16 \})$ complex,
\item Gauge: $2 N_d \times N_c^2$ complex.
\end{itemize}
Similarly we can count the floating point operations as proportional to the number of points in the stencil $N_{\rm stencil}$
and the number of fermion spin degrees of freedom per point after (any) spin projection $ N_{hs}$. The
cost of spin projection, and the cost of SU(3) matrix multiplication is
\begin{itemize}
\itemsep-0.3em 
\item  SU(3) MatVec: $66\times N_{\rm hs} \times N_{\rm stencil}$ .
\end{itemize}
We tabulate the arithmetic intensity and surface to volume ratio of the different kernels in Table~\ref{tab:arithint}.
\begin{table}[hbt]
\begin{tabular}{c|cccccccc}
Action  & Fermion Vol & Surface & $N_s$ & $N_{hs}$ & $N_{\rm rhs}$ & Flops & Bytes & Bytes/Flops\\
\hline
HISQ    & $L^4$ & $3\times 8\times L^3$ &  1  & 1 & 1 & 1146 & 1560 & 1.36  \\
Wilson  & $L^4$ & $8\times L^3$ &  4  & 2&  1 & 1320  & 1440 & 1.09      \\
DWF     & $L^4\times N$ & $8\times L^3$ &  4 & 2 & 16 & $N_{\rm rhs}\times 1320$ & $N_{\rm rhs}\times 864$ & 0.65\\
\hline
Wilson-RHS  & $L^4$ & $8\times L^3$ &  4  & 2 & 16 & $N_{\rm rhs} \times 1320$ & $N_{\rm rhs}\times 864$ &0.65 \\
HISQ-RHS    & $L^4$ & $3\times 8\times L^3$ &  1  & 1 & 16 & $N_{\rm rhs}\times 1146$ & $N_{\rm rhs}\times 408$& 0.36\\
\end{tabular}
\caption{\label{tab:arithint}
Arithmetic intensities and surface to volume ratio for different forms of QCD action and solver. Note that for Wilson and 
Staggered fermions the arithmetic intensity can be greatly improved for valence analysis by increasing the number 
of right hand sides. 
}
\end{table}
Since  $\sim \frac{1}{L}$ of Fermion data comes from off node, scaling fine operator requires interconnect bandwidth
$$B_{network} \sim \frac{B_{memory}}{L} \times R $$
where $R$ is the \emph{reuse} factor obtained for the stencil in caches, assuming that the memory bandwidth $B_{memory}$ is saturated.

\label{sec:netreq}
We use the above table of results to estimate the required interconnect bi-directional bandwidth per node to support
single node performance measurements for a variety of architectures, under the assumption that the communication time
should be equal to the computation time during the application of a fine Dirac operator, Table~\ref{tab:netreq}.
%
%
\begin{table}[hbt]
\begin{tabular}{|c|c|c|c|c|c|}
\hline
Nodes      & Memory (GB/s) & \multicolumn{4}{|c|}{Bidi network requirement (GB/s)}\\
           &         & L=10    & L=16 & L=32 & L=64 \\
\hline
2xBroadwell& 100 & 100 & 16 & 8 &\\ 
KNL        & 400 & 100 & 64 & 32&\\
P100       & 700 & 200 & 128 & 64&\\
DGX-1      & 5600& - & 975 & 487 & 243\\
\hline
\end{tabular}
\caption{\label{tab:netreq}
Estimate the required interconnect bi-directional bandwidth per node to support 
single node performance (as measured for a variety of architectures), for $L^4$ local volume 
assuming that the communication time should equal the computation time during the application of a fine Dirac operator. 
The assumption is particularly good if communications
can be efficiently overlapped with computation.
}
\end{table}

\section{Performance results}

We aggregate performance results from these new architectures provided by various members of the community.
Results on the older Knights Corner processor were reported by  Ishikawa-san\cite{ishikawa}, but we
focus on Knights Landing and Pascal results for this review.

{\bf Grid on KNL:}
Grid is an Edinburgh developed data parallel QCD package\cite{grid}.
It is designed to change data layouts to automatically adjust to the vector length 
of a given architecture, and delivers good performance (10-30\%) through the compiler
on a range of modern x86 architectures. The library also supports a architecture independent
target and the ARM and IBM BlueGene/Q instruction sets. Performance figures for 
the application of $L_s$ copies of the Wilson operator is provided in Table~\ref{tab:Grid}.

\begin{figure}[hbt]
\begin{minipage}{0.5\textwidth}
\includegraphics[width=\textwidth]{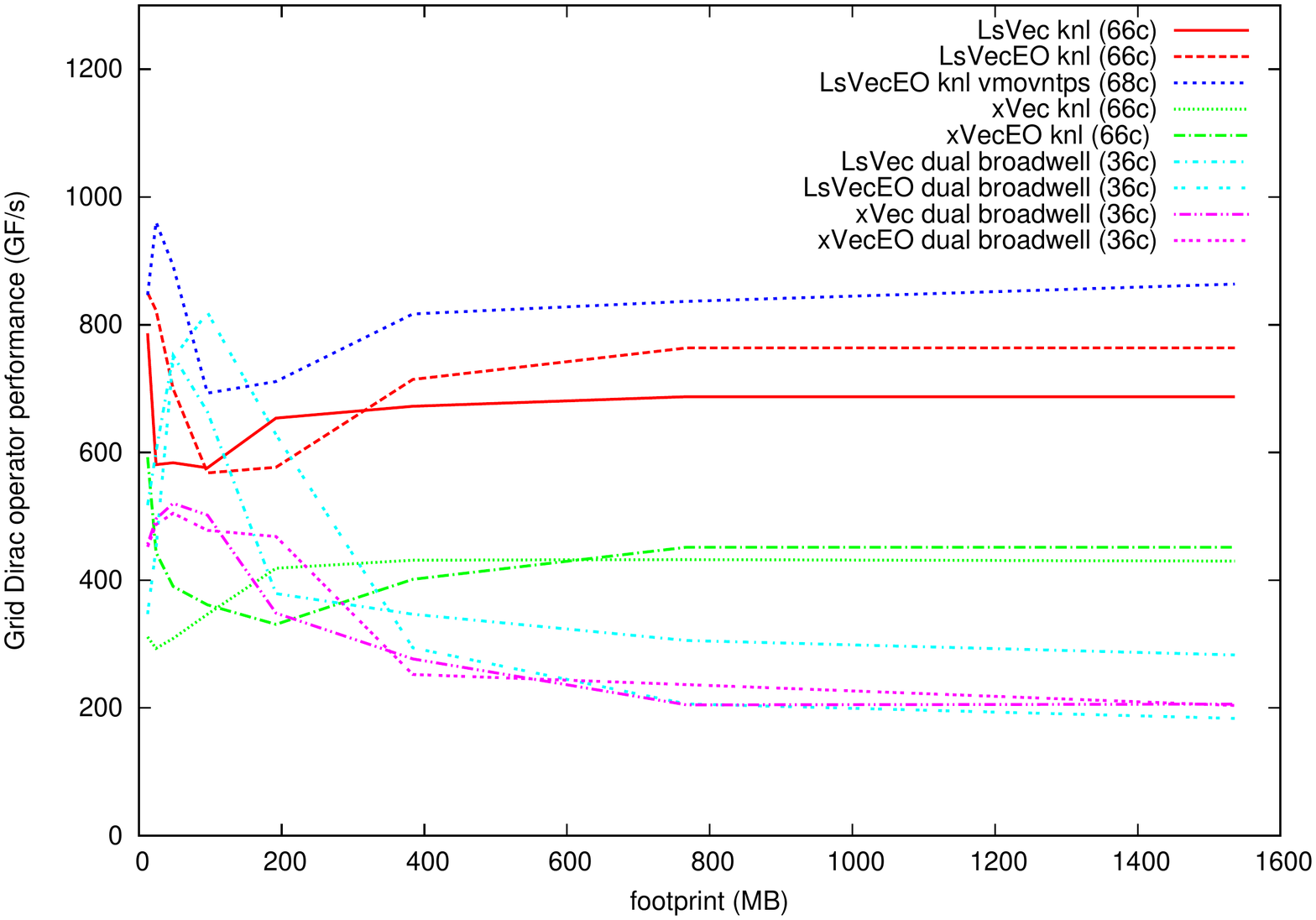}
\vspace{-0.05\textheight}
\end{minipage}
\begin{minipage}{0.5\textwidth}
\includegraphics[width=\textwidth]{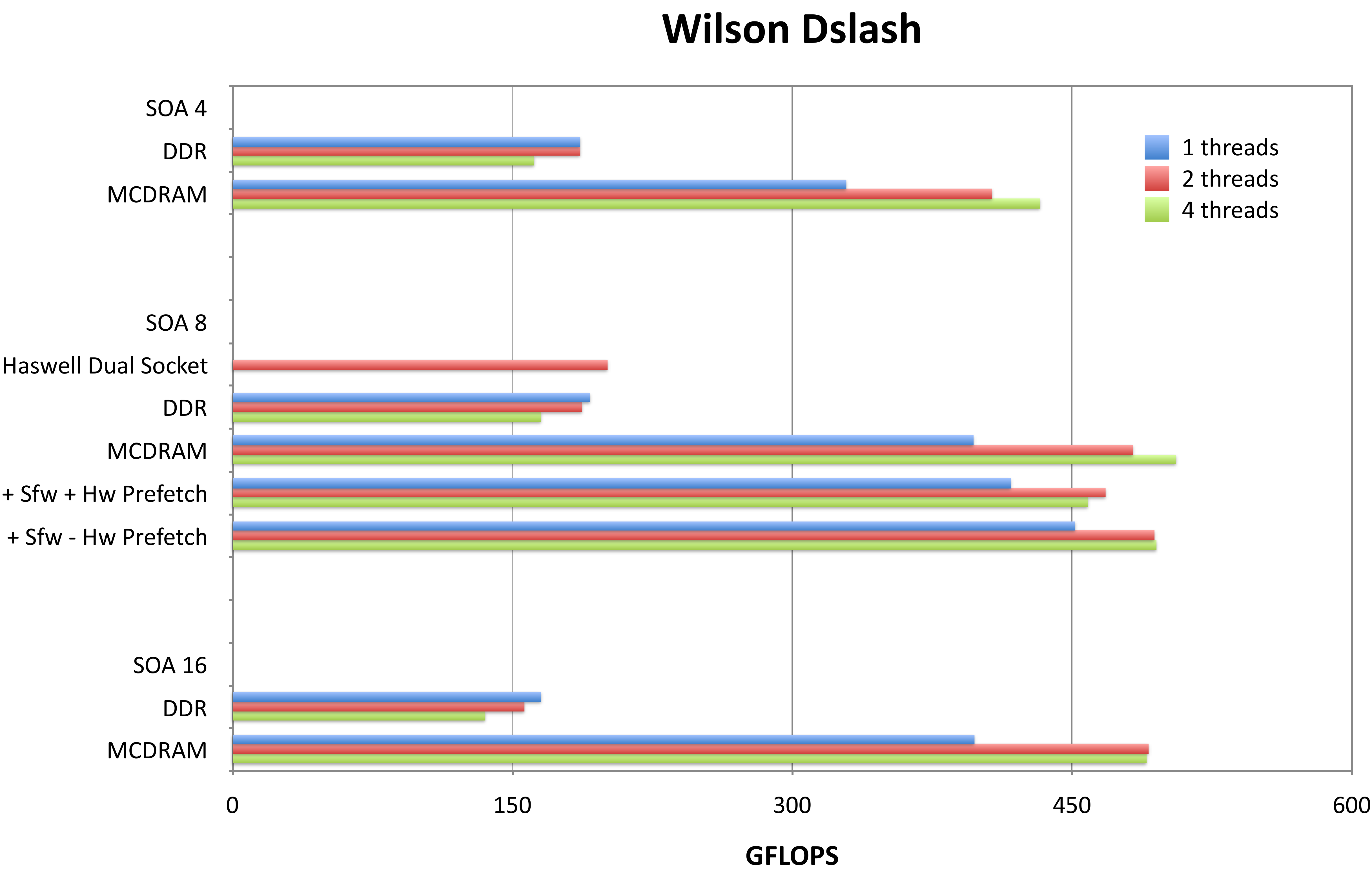}
\end{minipage}
\caption{
\label{fig:GridKNL} \label{fig:qphix1}
Left: Grid performance for $L_s=16$ applications of the Wilson operator on a single KNL 7250 node. The Haswell benchmarks
are from Cori Phase-1 32 core dual Haswell Cray XC40 nodes. The memory system performance of the KNL at large memory footprints 
demonstrates the efficacy of the memory integration, and the performance difference is marked. 
Right: The performance of the QPhiX Wilson dslash on KNL for a variety of partial vector layout transformations, hyperthreads per core, 
and running from both the 6 channel DDR memory and the on package MCDRAM. Again, the benefits of high performance 3D memory integration are
clear. In contrast to Grid assembly code, the best performance is obtain with more than one hyperthread per core since the latency tolerance
to stack evictions is greater.
}
\end{figure}

\begin{table}[hbt]
\begin{tabular}{cccc}
Architecture & Cores & GF/s (Ls x Dw) & peak\\
\hline
Intel Knight's Landing 7250   & 68 & 770 & 6100 \\
Intel Knight's Corner & 60 & 270  & 2400\\
\hline
Intel Broadwellx2 & 36 & 800  & 2700\\
Intel Haswellx2   & 32 & 640  & 2400\\
Intel Ivybridgex2 & 24 & 270 & 920\\
AMD Interlagosx4 & 32 (16) & 80 & 628\\
\hline
\end{tabular}
\caption{\label{tab:Grid} Performance of Grid across several architectures}
\end{table}
Grid has been particularly optimised for Knight's Landing as part of an Intel Parallel Computing
Centre at Edinburgh and Columbia, and the performance on a single KNL-7250 part is compared
to that of two Cori Phase-1 Haswell chips as a function of the local volume in Figure~\ref{fig:GridKNL}
The team report multinode single precision performance at 400GF/s on a $24^4$ local volume on an Omnipath network when
the job is spread out in four dimensions over $2^4$ nodes, after communications has been overlapped with computation.

One KNL was substantially faster than two Broadwell's (18+18 cores) when the footprint was larger than cache due to
the good performance of the high bandwidth on package memory.
For comparison the thermal design power (TDP)
for a dual Broadwell node is around 290W, while the TDP for a KNL node is around 210W, so that this increase in performance is accompanied by a
reduction in power consumption.
The Grid team report some other interesting metrics from performance measurements of their code:
\begin{itemize}
\itemsep-0.2em 
\item not using all the cores was fastest; typically two spare cores help performance
\item best performance using inline assembly (\emph{not compiler intrinsics}); this was due to the 
AVX512 instruction set having many registers and 3 operand instructions allowing hand allocation of registers
to excel. 
\item best performance was obtained with 1 thread per core on this highly optimised code; the L1 cache becomes
a deterministic state machine only when a single thread runs, and stack evictions are eliminated since reused data is 
protected in registers.
\item Single core instructions-per-cycle is 1.7 from a possible 2.0.
\item Multi-core L1 hit rate was 99\% due to perfect SFW prefetching
\item Multi-core MCDRAM read bandwidth saturated 97\% of streams (370GB/s)
\end{itemize}

\begin{figure}[hbt]
\includegraphics[width=\textwidth]{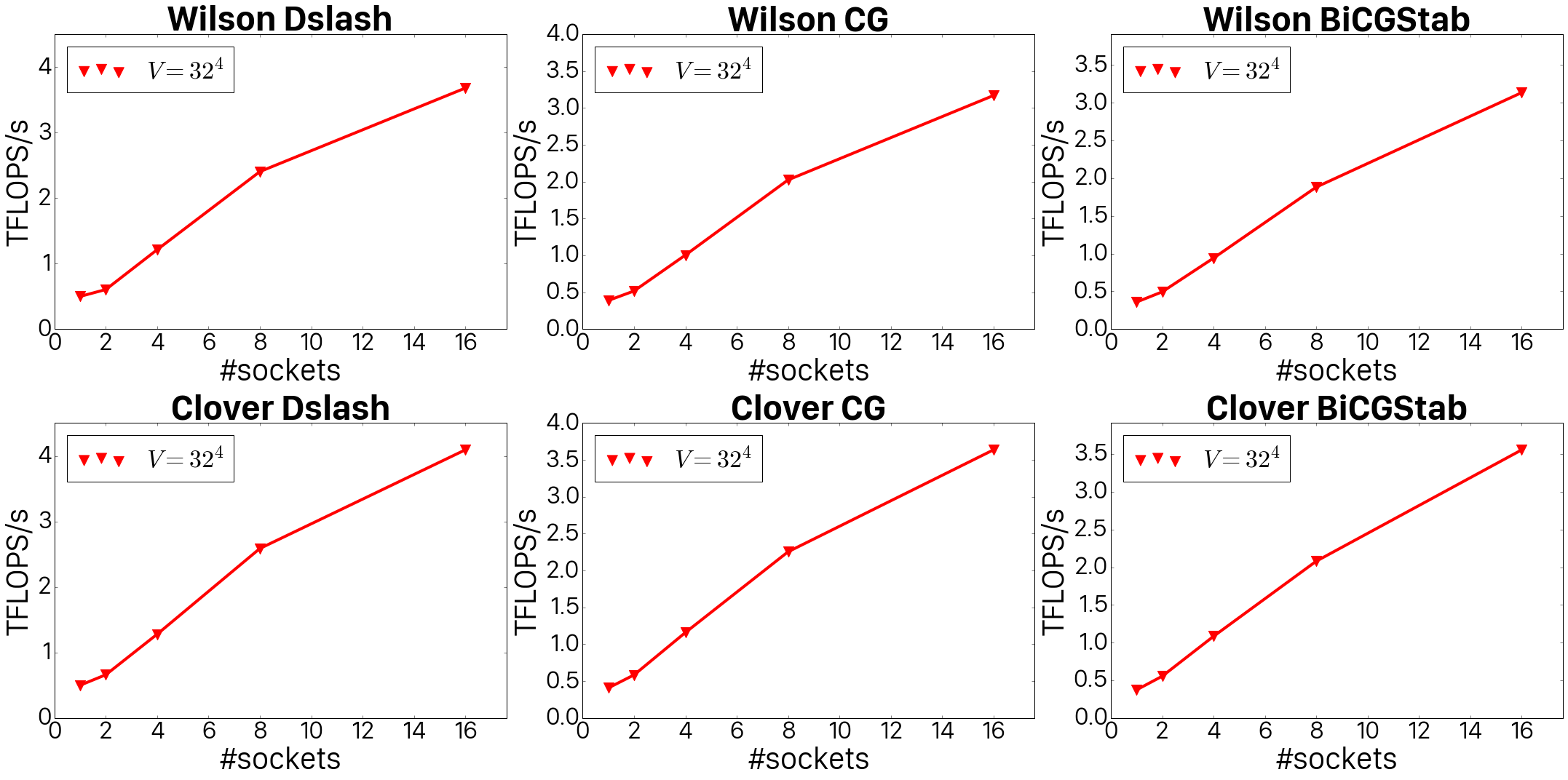}
\caption{\label{fig:qphix2} Multinode scaling of key QCD routines in KNL over a single rail Omnipath 1.0 network.
The Wilson and Clover Dslash routines give around 250GF/s per node in single precision in multinode code.}
\end{figure}

{\bf Wilson operator on KNL:}
The Intel and Jlab team developing QPhiX\cite{qphix} have ported their code to KNL and
provided both single node, Figure~\ref{fig:qphix1}, and multinode performance benchmarks, Figure~\ref{fig:qphix2}.
These results are in single precision and make use of a single rail Omnipath 1.0 network.
In contrast to the assembly programmed Grid benchmark, the intrinsics used rely on compiler 
register (non)allocation with the result that stack variables are (in practice) stochastically ejected from L1 cache by
streaming data. A consequence of this is that best performance is obtained with several threads per core.

\begin{figure}[hbt]
\includegraphics[width=0.3\textwidth]{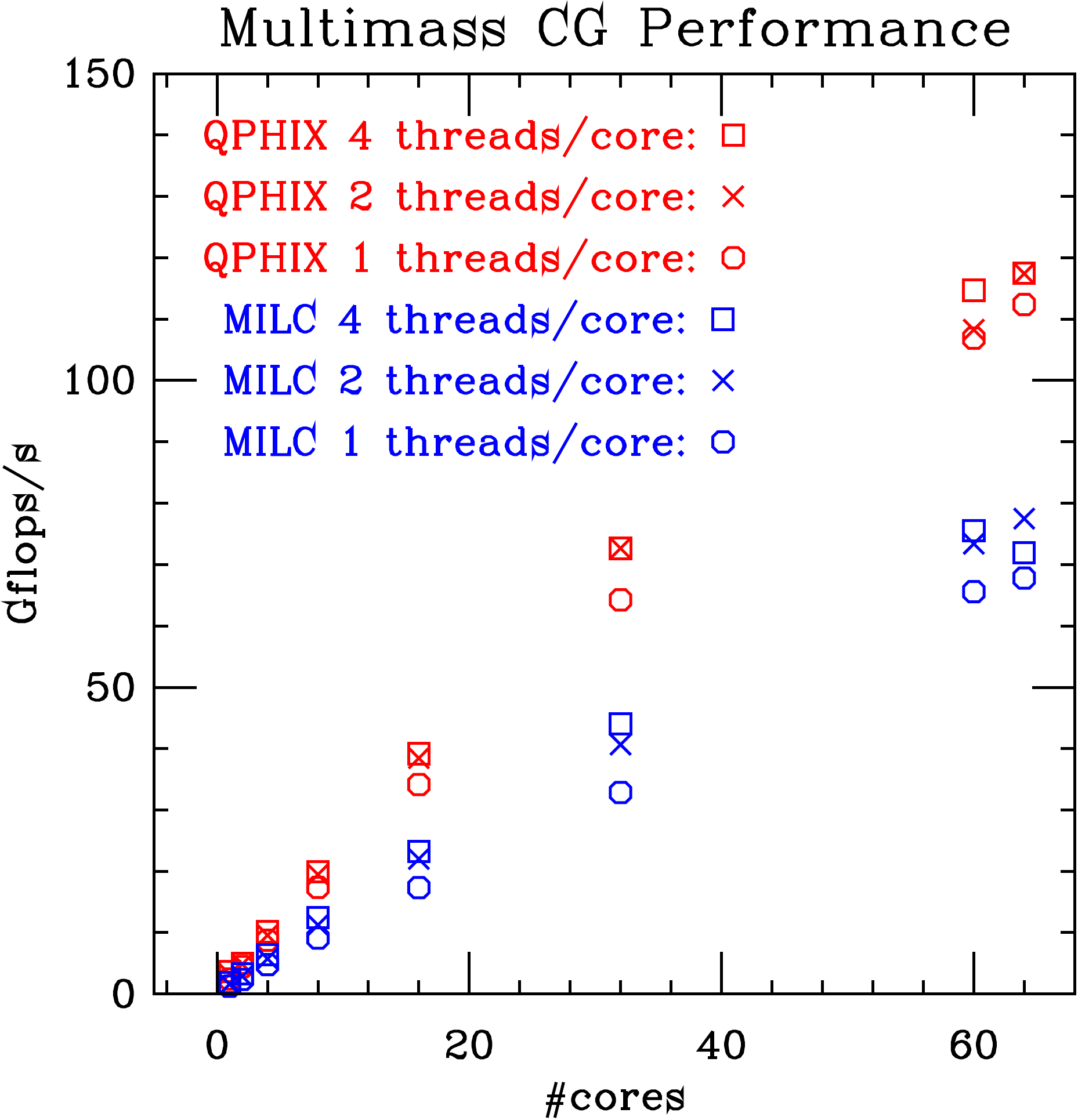}
\includegraphics[width=0.3\textwidth]{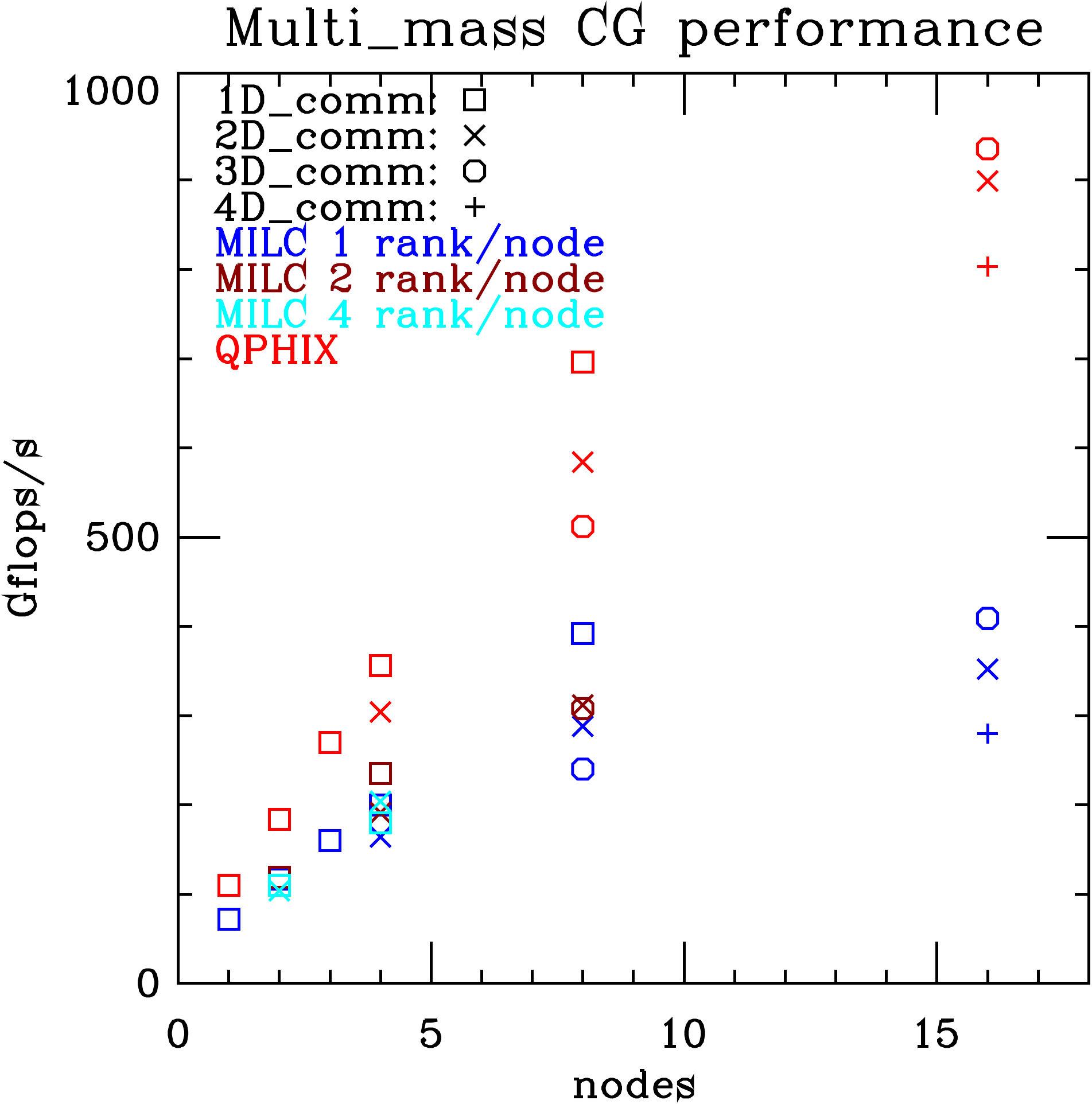}
\includegraphics[width=0.3\textwidth]{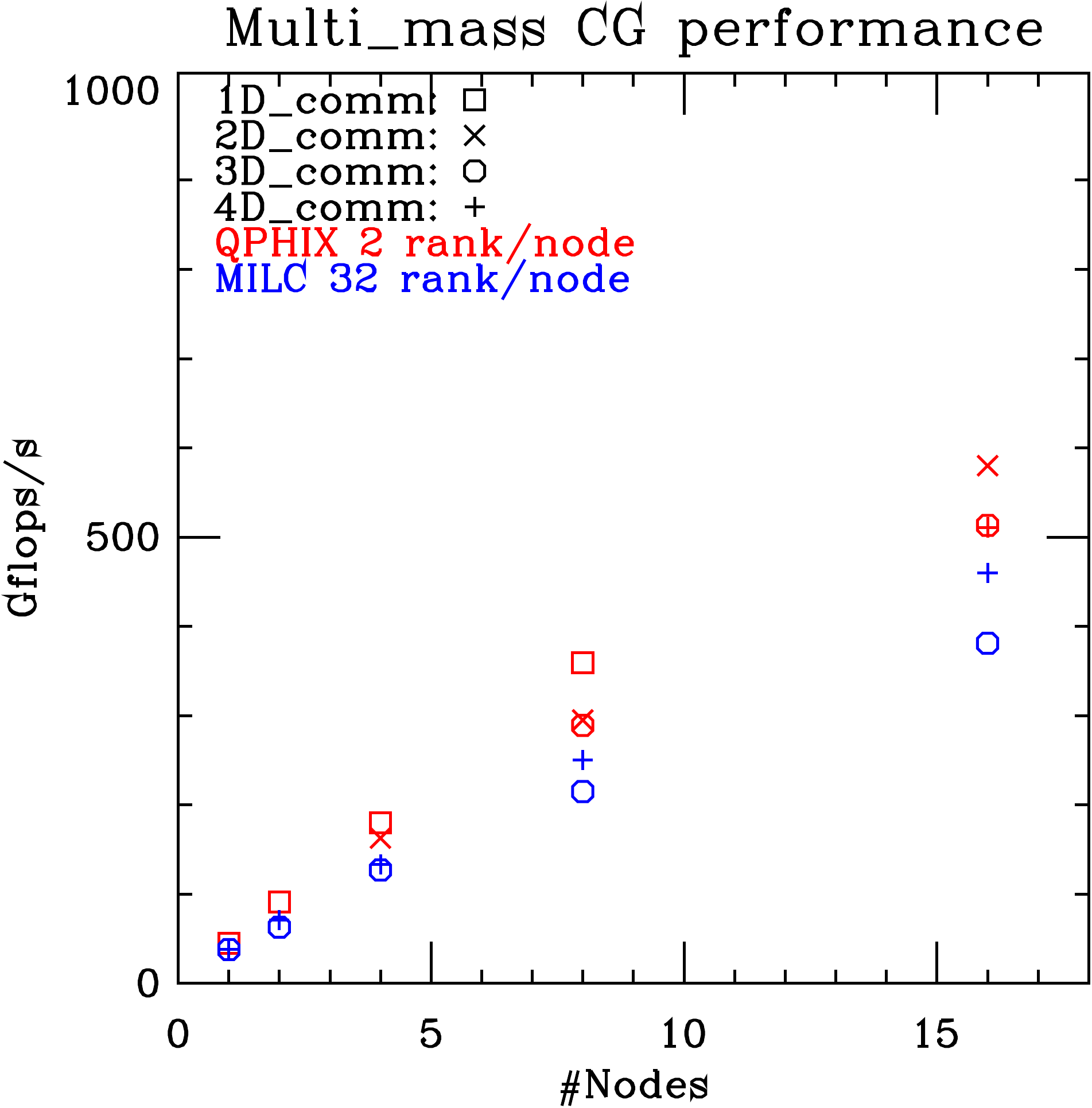}
\caption{\label{fig:milc}Left:
MILC code double precision single node multimass HISQ conjugate gradient performance in flat memory mode. The MILC+QPhiX code
substantially beats the scalar MILC code emphasizing the need to use compiler intrinsics or other vectorisation
schemes. The code is heavily linear algebra dominated (it is a multimass solver) and this limits performance. Around 80\% of
available memory bandwidth is consumed.
Right: Multinode KNL performance of the multimass solver. Since linear algebra dominates the communiction
the performance scales with node count very well, at least over the limited test range.
}
\end{figure}

\begin{figure}[hbt]
\includegraphics[width=0.5\textwidth]{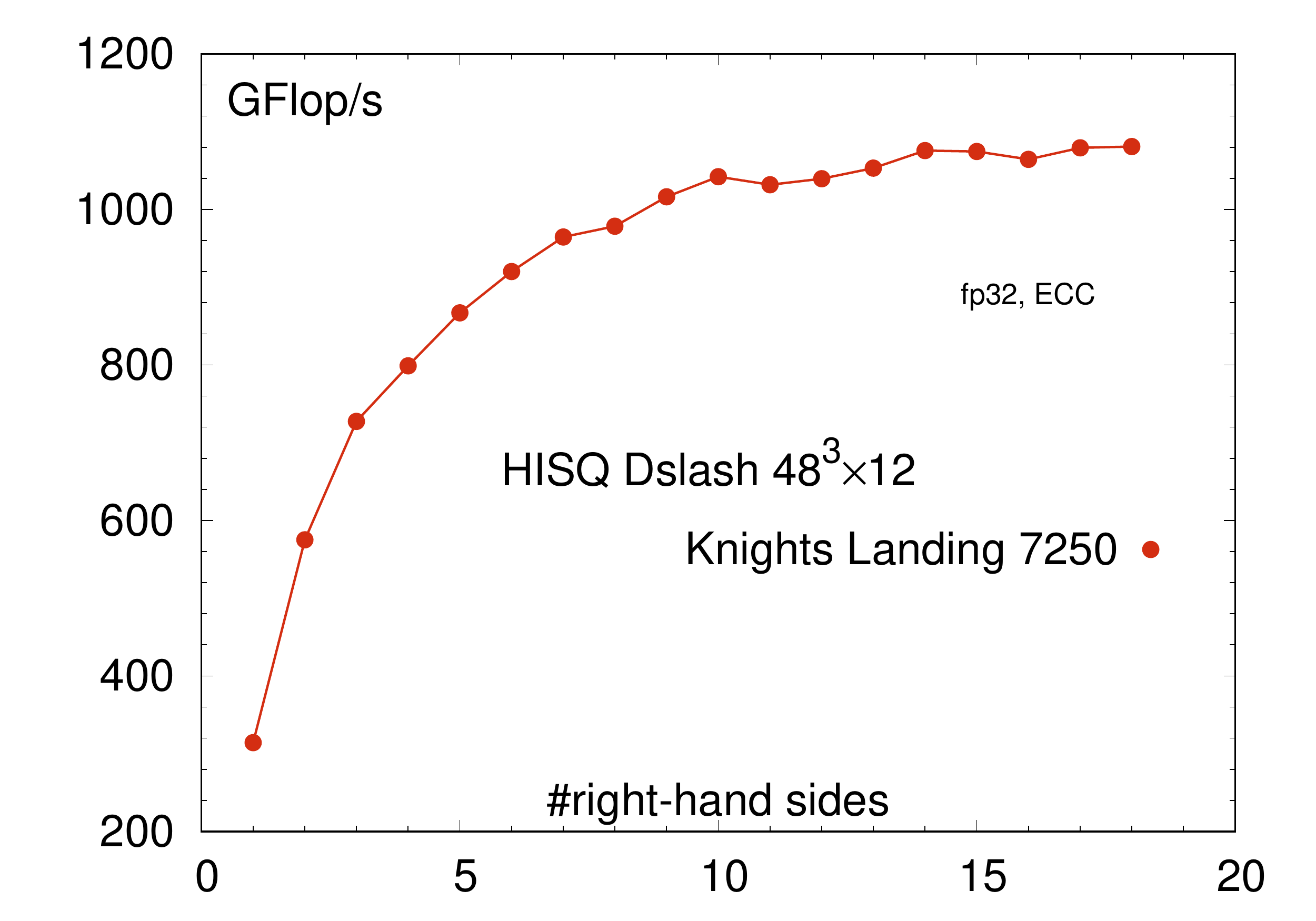}
\includegraphics[width=0.5\textwidth]{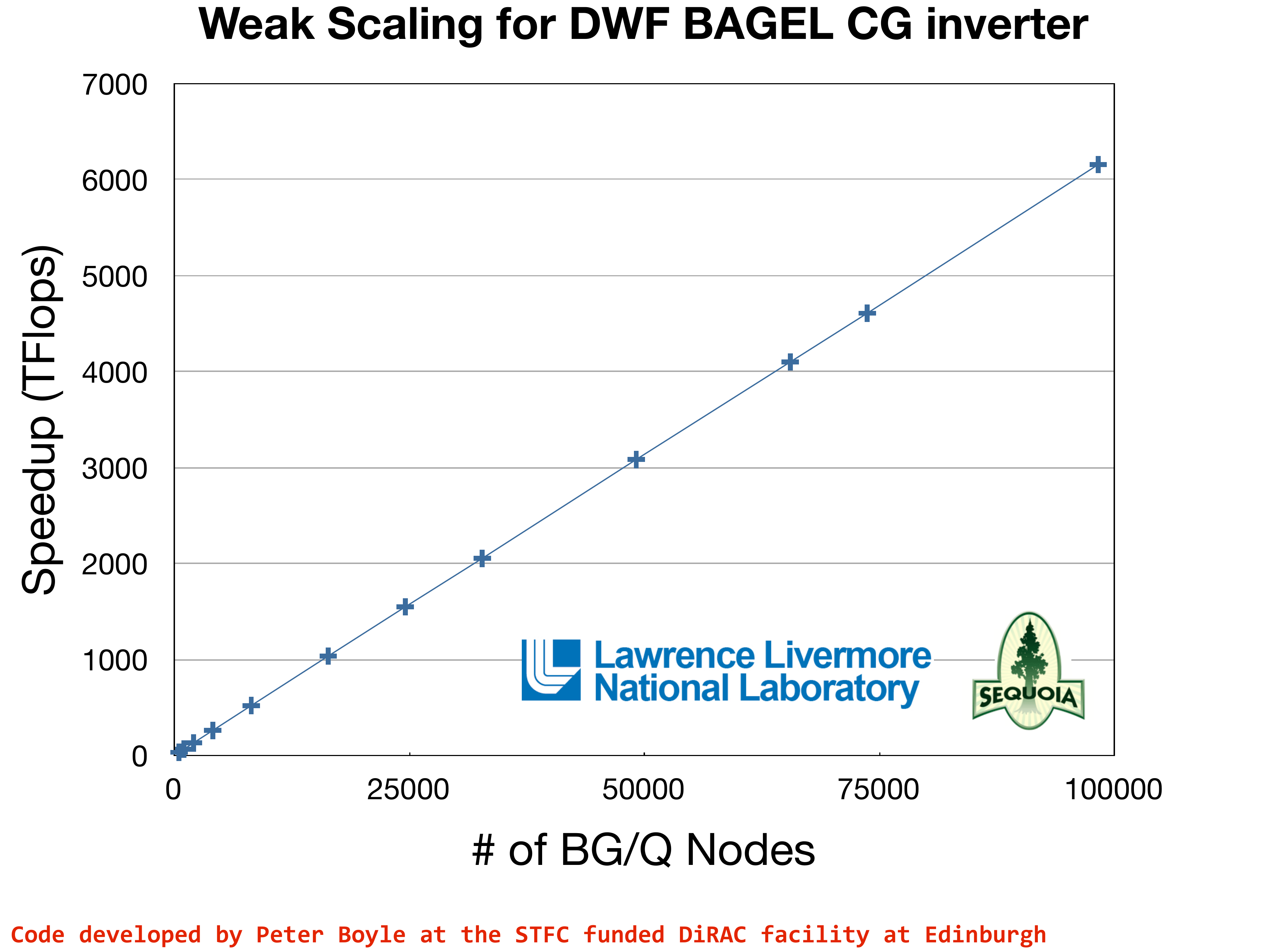}
\caption{\label{fig:patrick}
Left: Multinode Broadwell dual node performance of the multimass solver. Since linear algebra dominates the communiction
bandwidth that is common to both the Broadwell nodes and the KNL nodes is irrelevant in this benchmark and the memory bandwidth
of the KNL system leaves one KNL chip substantially faster than two Broadwell chips in a dual socket node.
Right: The single node HISQ Dslash developed by Patrick Steinbrecher (Bielefeld \& BNL)\cite{steinbrecher}
for multiple right hand sides saturates at around 1100 GF/s in single precision on a
KNL 7250 part. When combined with the algorithmic efficiency of block solvers\cite{Wagner}, this
approach has multiplied attractions of giving greater numerical and algorithmic performance and is compelling.
Right: A reminder of how far scaling can be pursued (to 1.6M cores) on BlueGene/Q where the integration of 
strong interconnect bandwidth in a way that scales well with the floating point performance enables
a cheap route to massive scalability. In contrast, if powerful nodes are coupled to weak networks most
of the available floating point would lie idle if even moderate node counts were used on on a tractable lattice volume.
There is no substitute for around 1 GB/s per sustained Gflop/s.
}
\end{figure}

{\bf Staggered Fermion performance on KNL:}
The MILC multi-mass CG was presented by Ruizi Li, Figure~\ref{fig:milc} and Figure~\ref{fig:patrick}.
This was based on a multimass solver in double precision and is heavily memory bandwidth limited. Indeed
interconnection performance for this algorithm (depending on the number of masses, and the degree to which
they converge after similar iteration counts) is somewhat irrelevant. In contrast, the multiple
right hand side approach taken by Steinbrecher, Figure~\ref{fig:patrick} is much more cache friendly and
performs better. This latter approach is most useful for valence analysis. Steinbrecher's code 
does not use MPI and is single node only, and is applied for trivially parallel disconnected diagram analyses.
A multinode version of the same code would be rather more senstive to interconnect bandwidth than the benchmark
discussed by MILC.

\begin{figure}[hbt]
\includegraphics[width=0.5\textwidth]{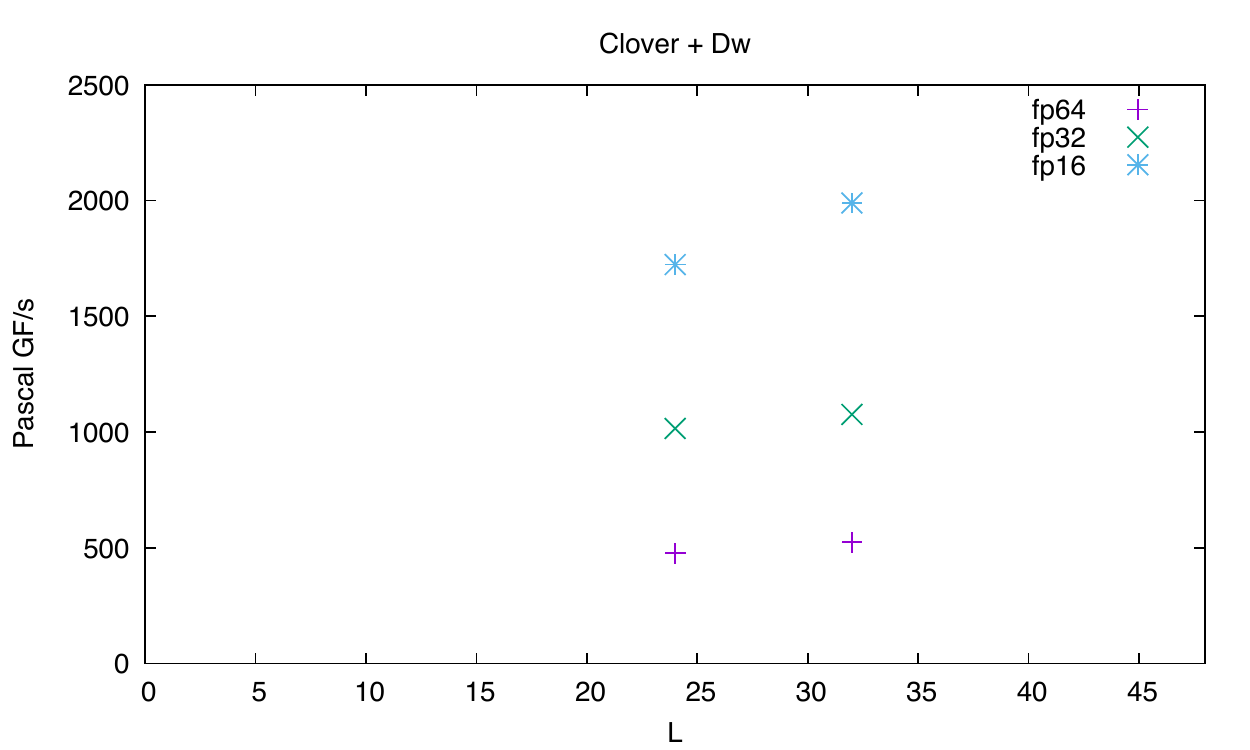}
\includegraphics[width=0.5\textwidth]{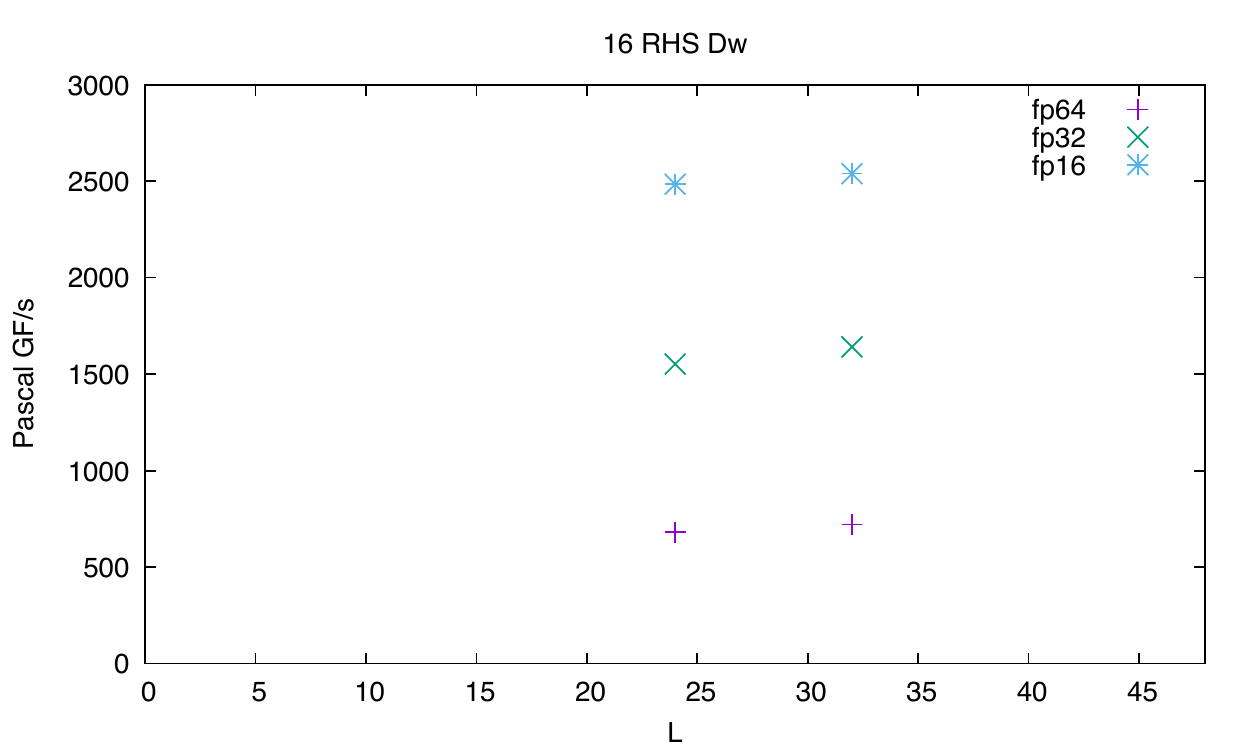}
\caption{\label{fig:nvidia1}
Left: Clover term + $D_W$ relevant to the solution of standard Clover-Wilson fermions on the Pascal GP100
GPU in single precisoin.
Right: 16 RHS $D_W$ relevant to multiple Wilson RHS solvers and to various 5d Chiral fermion approaches
such as DWF. Results are in single precision on a Pascal GP100. 
}
\end{figure}

\begin{figure}[hbt]
\includegraphics[width=0.5\textwidth]{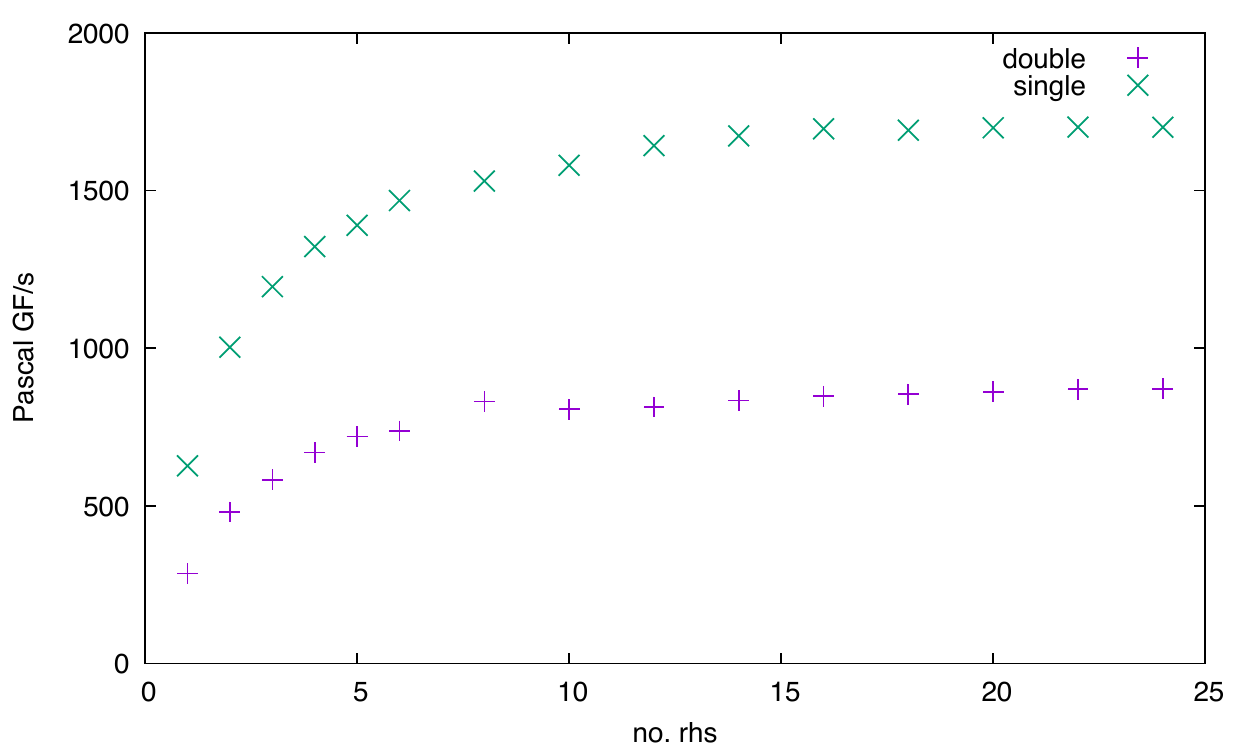}
\includegraphics[width=0.5\textwidth]{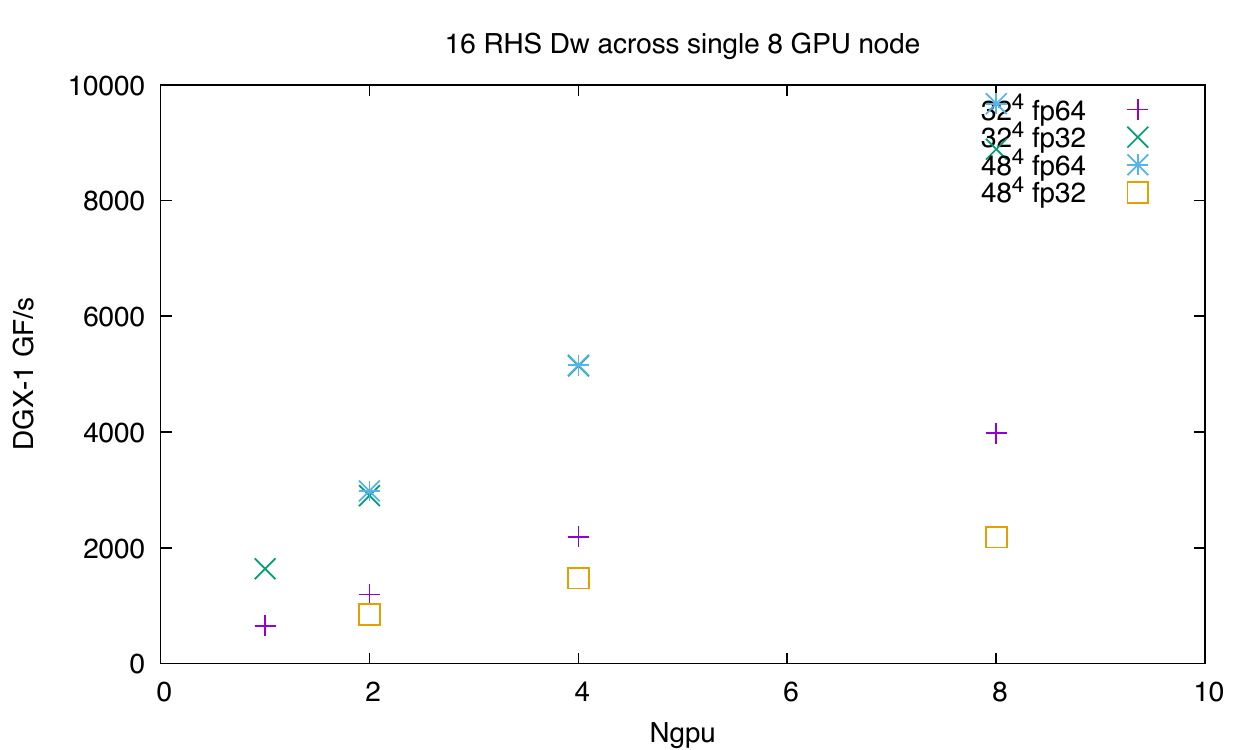}
\caption{\label{fig:nvidia2}
Left: Multi RHS Staggered matrix multiply on the GP100 part. Again roughly 2x performance per node is
obtained compared to KNL in single node code.
Right: Scaling across DGX-1 8 gpu system; for sufficiently large volumes (e.g. $48^4$) the scaling is linear
to 8 nodes, and around 10TFlop/s.
}
\end{figure}

{\bf Nvidia Pascal Results:}
We focus first of all on results produced by Nvidia with the QUDA library.
In these nodes a single GPU is capable of 5.4/10.6/21.2 TF/s in double/single/half
precisions respectively.
The Pascal results and code were provided by Kate Clark, Nvidia, and these
comprise (a) the Wilson kernel and it cache friendly application to 5d chiral
fermions (or multiple RHS solvers) Figure~\ref{fig:nvidia1} and (b) the 
multiple right hand side improved Staggered fermion matrix and the scaling across
an 8 GPU DGX-1 node Figure~\ref{fig:nvidia2}. 
In all cases the single node figures are roughly double a Kights Landing processor for single node execution,
and this is consitent with the substantially improved HBM2 memory bandwidth compared to KNL. 
Multinode code will depend critically on the interconnect performance, and as indicated in Section~\ref{sec:netreq}
it is unlikely to be able to provision sufficient interconnect to produce a scalable system that does not
have the local compute outstrip the network. It remains possible that multigrid algorithms with sufficiently
aggressive domain decomposition preconditioners may scale.
For conventional Krylov solvers, 
the ability of these nodes to scale across an MPI network is likely compromised, but at least for valence analysis
multiple 10TF/s sustained nodes are not an unreasonable proposition. 
%
Unfortunately, the DGX-1 costs roughly \$130,000 list price\cite{next}
or an eye watering \$ 16,000 per GPU. The canny academic could no doubt maintain a fall
back plan of using commodity parts and holding the pricing rather closer that of similar parts (e.g. Titan-X)
sold to the volume gaming market. For example,
an aggressive research group could tolerate less than 100\% reliability by verfiying solutions
to the Dirac equation with retry loop, or may even negotiate the better parts at a more reasonable price.


Offload to GPU poses difficulty to  code maintainance and performance portability, and there
has been a big community investment in QDP-JIT, for example, and vendor investment in 
architecture specific libraries such as QUDA.
It was reported in this conference that the author and  Meifeng Lin reduced the Grid expression
template engine to 200 line-of-code example, and succeeded  to offload with a CUDA kernel call 
to evaluate expression using ``compile time compilation'', and specifically withour resorting to Just-In-Time
compilation. This task will become even easier with unified memory model introduced with Pascal GPU's.

James Osborn presented a package called ``Quantum Expressions'', QEX, based on the ``NIM'' language\cite{nim}.
The language, although not widespread or mature, has many attractive features and enables a controllable 
``nim to C'' mapping. This in principle enables more flexible translation to GPU and indeed other targets,
with more easily controllable mapping than is afforded by C++ template metaprogramming. Osborn believes this
can generate equivalent or greater performance to C++ packages with greater flexibility and portability.

\begin{figure}[hbt]
\includegraphics[width=0.5\textwidth]{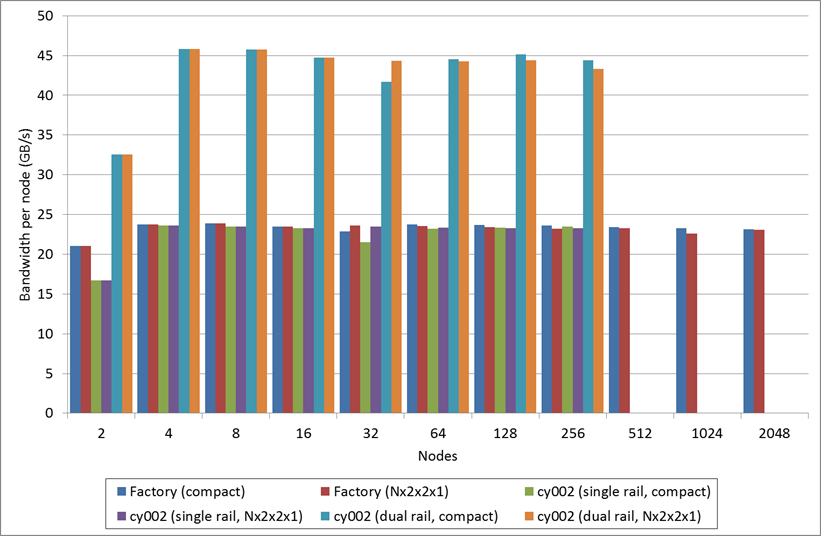}
\includegraphics[width=0.5\textwidth]{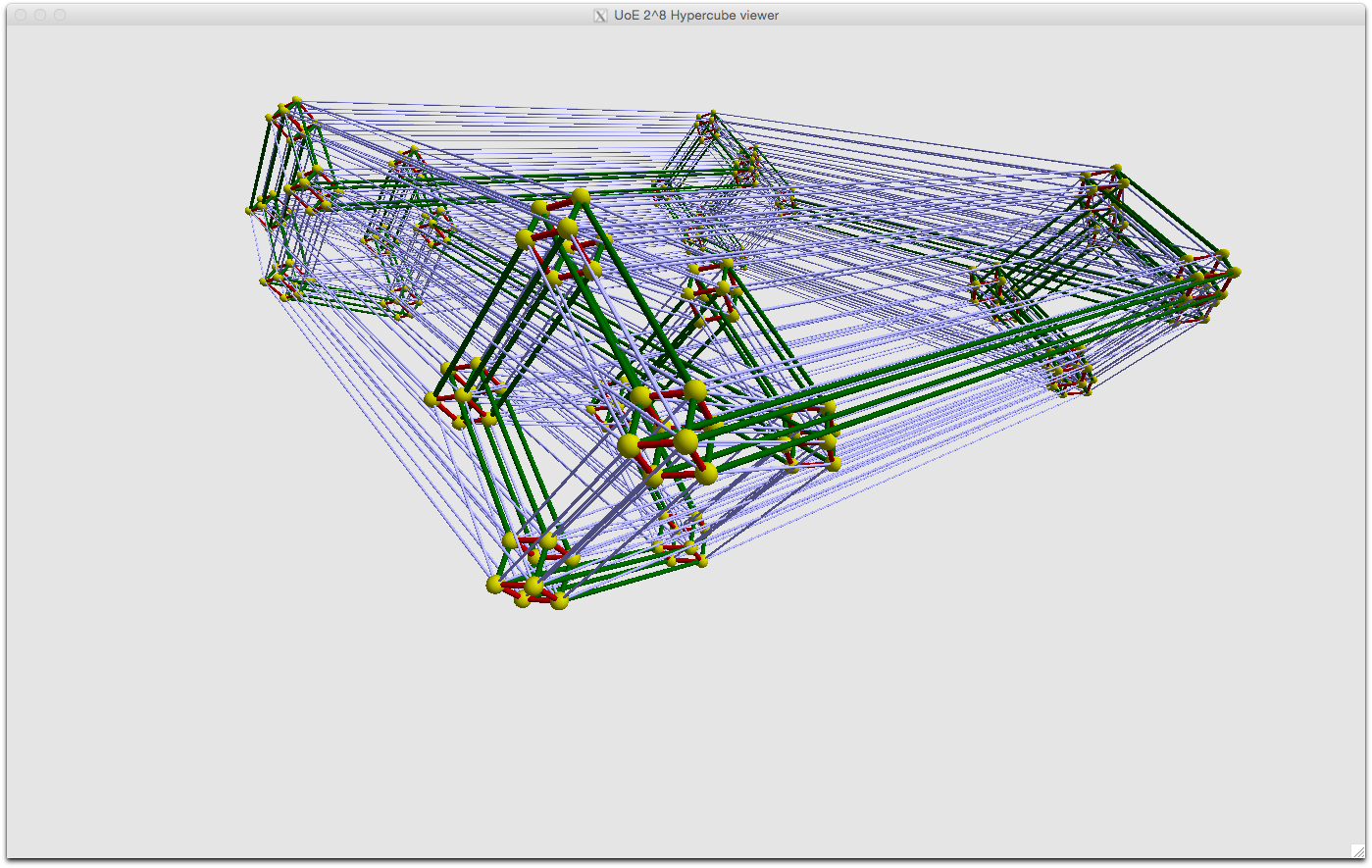}
\caption{\label{fig:bgqncar}
We display results (left) from SGI having developed a perfect mapping scheme between logical and physical
interconnect topologies(right). No weak scaling slow down is seen even for the largest system sizes. Results
were produced on the Cheyenne system at NCAR. Compared to the default MPI Cartesian communicator a 4x speedup
is seen on the largest partition sizes, demonstrating the effectiveness of topology aware job placement.}
\end{figure}

{\bf Communications performance results:}
As has been discussed, \ref{sec:netreq}, performance in a massively parallel calculation is critically dependent
on interconnect bandwidth. The Grid benchmark suite has been used to characterise many of the popular interconnects
in forthcoming machines, Table~\ref{tab:network}. 
We see that compared to the BlueGene/Q 32 GB/s sustained bidirectional
bandwidth in 2012, there is suprisingly little advance four years later\cite{boylebgq}. With a 205GF/s peak
node, the BlueGene/Q enable QCD scalability to 1.6 million cores, delivering 7.2 PF/s, Figure~\ref{fig:bgqncar}. 
This dwarves the figures
of 10TF/s sustained scale figures beyond which we might fear that Dirac matrix applicatoin will not
scale on a 8 GPU node. We can see that in principle we should seek to add network bandwidth in a way that
scales with the node count, and hence the importance of integration of the interconnect.
In this respect, the KNL-F design and the subsequent KNH remain interesting as our
figures demonstrate the BlueGene/Q was scalable to system sizes beyond those the QCD community can actually afford.

Since network becomes the large system performance limitation, we might seek to minimise network
cost while maintaining peformance for QCD calculation. Both the QCDOC design\cite{QCDSP}, and subsequently
QCD codes on the K-computer \cite{tofu} and BlueGene/Q\cite{boylebgq} have made use of precise mapping of
the application logical 4d torus to underlying interconnect topologies.
We present results provided by the Grid team in collaboration with Silicon Graphics, developing this theme
on the ICE-X hypercube. SGI ran benchmarks on custom partitions of their machines with an exact power of two
set of nodes on each switch (not the default) and demonstrated a 4x improvement in QCD scalability on 2048 node
partitions. Further, when using the Mellanox EDR interconnect techonlogy, 80\% of dual rail bidirectional
bandwidth was obtained with no performance loss as weak scaled to any system size. This is a similar characteristic
to the scaling of torus application codes on torii, although the job placement mapping was rather complex, Figure~\ref{fig:bgqncar}.
With the correct MPI mapping, one can embed the QCD torus inside hypercube so that nearest neigbour comms travel only a single hop.

\begin{table}[hbt]
\begin{tabular}{|c|c|c|c|c|c|c|}
\hline
Machine & Node & Network      &  1 rank/node & 4 ranks/node & Peak  & Require\\
\hline
Cori-1 & KNL &  Cray Aries  & 11 GB/s    & -        & - & 64\\
SGI    & KNL & Single EDR   & 23 GB/s    & -        & 25GB/s& 64\\
SGI    & KNL &   Dual EDR   & 45 GB/s    & -        & 50GB/s& 64\\
Brookhaven & KNL & Dual Omnipath& 14 GB/s    & 44 GB/s  & 50GB/s& 64\\
\hline
\end{tabular}
\caption{\label{tab:network}
Delivered measurements of network bandwidth from Grid communications benchmark}
\end{table}

\section{Algorithms}

I have chosen to focus on two aspects that I feel are most fundamental to continued progress.
These are multi-scale fermion solvers and block solvers, and multi-scale integration.
I will not cover in detail, even though they are also fundamental,
topological sampling (covered by Michael Endres' review) and specifically metadynamics
\cite{sanfilippo}. 
I pause only to comment on both that symptomatic relief is not necessarily a cure;
We ideally want solutions that address all forms of critical slowing down in an exact MCMC chain that is run
far enough to converge on the fixed point of the process. Fast thermalisation or creating a distaste for zero
topological does not necessarily accelerate ergodic sampling of the fixed point distribution which might
become slow even within a topological sector.
I also do not review approaches to free energy, density of states and derived observables, as these were reviewed by Langfeld
\cite{Langfeld,Nada,Pellegrini, Lucini}.

\begin{figure}[hbt]
\includegraphics[width=0.49\textwidth]{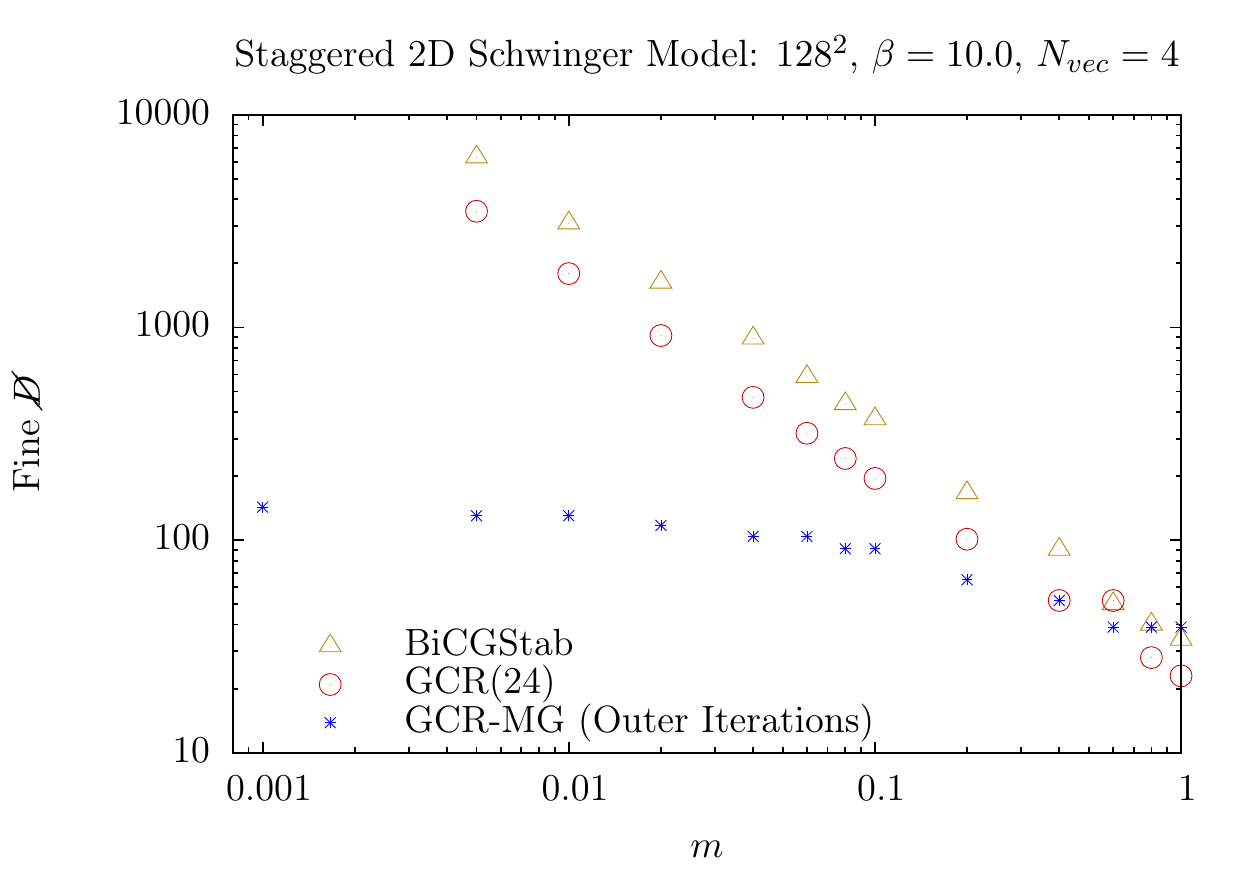}
\includegraphics[width=0.45\textwidth]{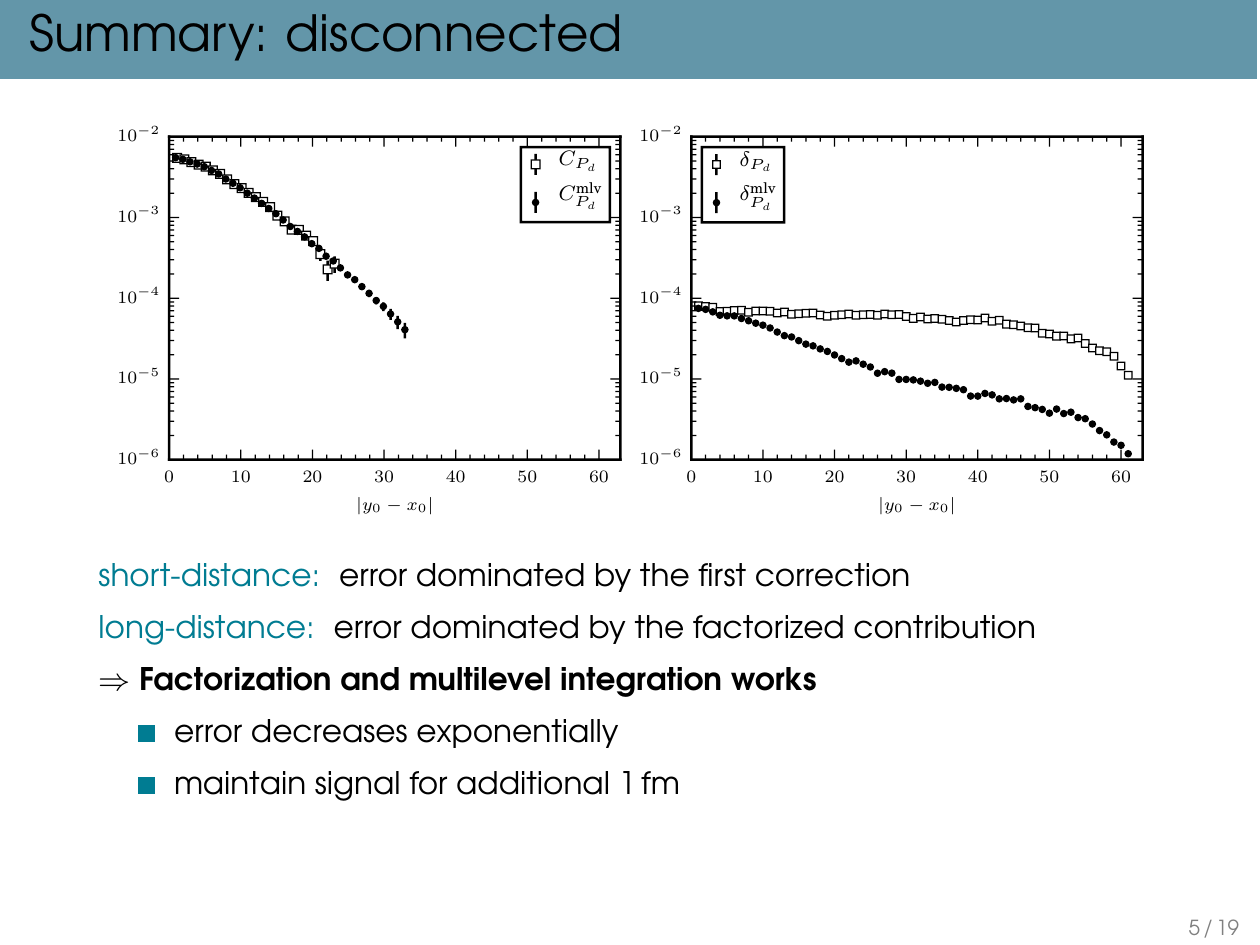}
\caption{\label{fig:mgrid} Left: 2D staggered multigrid elimination of critical slowing down.
Right: Improvement in a disconnected correlation function from multi-level integration.}
\end{figure}

{\bf Multiscale Fermion solvers:}
Several advances were reported in the area of Fermionic multigrid.
Weinberg, Brower, Clark, Strelchenko reported on steps towards a multigrid
solver for staggered fermions with demonstrated elimination of critical slowing down
in two dimensional systems\cite{weinberg}, Figure~\ref{fig:mgrid}. Yamaguchi and Boyle\cite{yamaguchi}
demonstrated the first solver and multigrid algorithm to work with 5D chiral fermions without
squaring the operator, by using the $\Gamma_5$ hermitian operator. This is a key step to keep the compact radius of the coarse space operator
limited and should enable a true, recursive multigrid. The authors claim to have understood the
connection between the spectrum of the 5d operator and the constraints on which Krylov solvers work,
and they hope that the nearest neighbour coarse space
operator will be sufficiently cheap to recompute that the acceleration can be applied in HMC evolution.

Stefan Schaefer and Marco Ce presented a promising quenched algorithm for multilevel integration\cite{schaefer}.
By subdividing the lattice into $m$ disjoint regions with boundaries which are \emph{independently} integrated
over, they were able vastly improve the statistical samples by a factor $N^m$ where $N$ represents the samples
drawn within each sub-region. For disconnected correlation functions this was demonstrated to give a substantial
win, Figure~\ref{fig:mgrid}. Further, elements of the DD-HMC determinant factorisation appear to be reusable
making it fairly likely that these authors will be able to pursue this algorithm to a viable approach for
dynamical fermions. Consequently, this is perhaps the uniquely most exciting algorithmic development of the conference.
Further multigrid works include the development of a twisted mass multigrid, and the
DD-$\alpha$-AMG solver library \cite{Bacchio}. The implementation of TWQCD's Exact one flavour algorithm for DWF
was discussed\cite{murphy}.

\section{Summary}
There has been a tremendous growth in computer power from many core CPU's and GPU's, with this
years new products including Intel's Knights Landing giving 0.5-1TF/s single node single precision
sustained performance, and Nvidia's Pascal giving 1-2 TF/s single node single precision sustained performance.
However, interconnects are not keeping pace; this is disappointing since integration of interconnect
in a manner that grows with the node transistor count should lead to scalable computing. The integration
of interconnect on KNL-F and Skylake nodes is interesting but significant work may still be required to obtain
scaling of QCD.
The use of fp64/fp32/fp16 arithmetic in preconditioners or variance reduction is not yet fully explored, and could
vastly both accelerate node computation, and in a network limited algorithm could be used in intelligent algorithms
or rounding robust to minimise network overheads.

Multigrid solver algorithms have recently solved critical slowing down in valence sector for Wilson/Clover,
and in this conference new multigrid algorithms have appeared for other actions (Staggered, DWF, Twisted Mass)
Successful application of multigrid in HMC exists for Wilson/Clover, but is not yet widespread.
Multilevel integration algorithms introduced in this conference are deeply interesting.
Algorithms that maintain ergodicity are a big challenge to using this power usefully, but were reviewed elsewhere in
the confernce.

{\bf Acknowledgements:} The author is an Alan Turing Institute Faculty Fellow, acknowledges 
STFC grants ST/M006530/1, ST/L000458/1, ST/K005790/1, ST/K000411/1, ST/H008845/1, and ST/K005804/1, 
and Royal Society Wolfson Research Merit Award WM160035.

\end{document}